\documentclass[12pt]{iopart}
\expandafter\let\csname equation*\endcsname\relax
\expandafter\let\csname endequation*\endcsname\relax
\usepackage{amssymb}
\usepackage[dvips]{graphicx}
\usepackage{enumerate}
\usepackage{epsfig} 
\usepackage{subfigure}
\usepackage{xcolor}
\usepackage[T1]{fontenc}
\usepackage{fullpage}
\usepackage{amsthm,amsfonts,amssymb,amscd,mathrsfs,xspace,framed}
\usepackage{mathrsfs,amsmath}
\usepackage{color}
\usepackage{setspace}
\usepackage{url}
\usepackage{wrapfig}
\usepackage{tikz}
\usepackage{enumitem}
\usepackage{bm}
\usepackage{comment}
\usepackage{txfonts}
\usepackage{array}
\usepackage{color}
\usepackage{braket}
\usepackage{physics}
\usepackage{graphicx}
\usepackage[english]{babel}
\usepackage{url}
\usepackage{bm}
\usepackage{hyperref}
\definecolor{darkblue}{rgb}{0,0,0.5}
\hypersetup{
colorlinks=true,
linkcolor=black,
filecolor=blue,
citecolor=darkblue,  
urlcolor=black,
}

\usepackage[square,sort&compress,numbers]{natbib}
\usepackage{cleveref}

\newcommand{\bp}{\boldsymbol \rho}
\newcommand{\bn}{\boldsymbol \eta}

\newcommand{\calN}{{\cal N}}

\def\be{\begin{equation}}
\def\ee{\end{equation}}
\def\ba{\begin{eqnarray}}
\def\ea{\end{eqnarray}}

\begin{document}

\title{Distributed Quantum Sensing}

\author{Zheshen Zhang$^{1,2}$, Quntao Zhuang$^{2,3}$}

\address{
$^1$Department of Materials Science and Engineering, University of Arizona, Tucson, Arizona 85721, USA
}
\address{
$^2$James C. Wyant College of Optical Sciences, University of Arizona, Tucson, AZ 85721, USA
}
\address{
$^3$Department of Electrical and Computer Engineering, University of Arizona, Tucson, Arizona 85721, USA
}
\ead{zsz@arizona.edu}
\ead{zhuangquntao@email.arizona.edu}

\begin{abstract}
A plethora of applications hinge on a network or an array of sensors to undertake measurement tasks. A rule of thumb for sensing is that a collective measurement taken by $M$ independent sensors can improve the sensitivity by $1/\sqrt{M}$, known as the standard quantum limit (SQL). Quantum resources such as entanglement and squeezed light can be harnessed to surpass the SQL. Distributed quantum sensing is an emerging subject dedicated to investigating the performance gain enabled by entangled states shared by multiple sensors in tackling different measurement problems. This Review formulates distributed quantum sensing from a quantum-information perspective and describes distributed quantum sensing protocols and their experimental demonstrations. The applications enabled by distributed quantum sensing and an outlook for future opportunities will also be discussed.

\end{abstract}

\tableofcontents

\maketitle
\section{Introduction}

The overarching goal of quantum information science and its enabled emerging quantum technologies is to demonstrate capabilities beyond what is allowed by their classical counterparts. Since the discovery of Shor's factoring algorithm~\cite{Shor_1997}, quantum computing has received significant interests and investments as it is widely deemed a disruptive paradigm capable of solving many classically intractable optimization~\cite{farhi2014quantum,steffen2003experimental,caneva2011chopped,venturelli2015quantum,moll2018quantum}, artificial intelligence~\cite{schuld2019quantum,lloyd2013quantum,wichert2013principles,dunjko2016quantum,biamonte2017quantum,dunjko2018machine,faber2018alchemical}, and big-data analysis problems~\cite{wiebe2012quantum,weinstein2013analyzing,rebentrost2014quantum,li2015experimental}. A remarkable milestone for quantum computing is the accomplishment of ``quantum supremacy'' on Google's 53-qubit superconducting quantum computer in solving a quantum random sampling problem~\cite{arute2019quantum}. Despite the tremendous recent advances in quantum computing, immense challenges arise when utilizing the available noisy intermediate-scale quantum (NISQ) hardware~\cite{Preskill2018quantumcomputingin} to implement practically useful quantum algorithms due to a lack of fault-tolerant mechanism. A continued pursuit for the quantum-information community is to seek application scenarios in which quantum technologies enjoy a quantifiable advantage over their classical counterparts.

Sensing is an arena that quantum technologies can achieve advantages over classical sensing technologies for practical applications in the near term. Quantum metrology~\cite{giovannetti2006,giovannetti2011advances} studies the use of nonclassical resources to enhance the performance of measurements for a variety of sensing applications. As a prominent example, the Laser Interferometer Gravitational-wave Observatory (LIGO) injects nonclassical squeezed light into its Michelson interferometer to surpass the standard quantum limit (SQL) due to laser shot noise~\cite{abadie2011gravitational,aasi2013enhanced,tse2019quantum}. Apart from LIGO, quantum metrology has also been exploited in target detection~\cite{Lopaeva_2013,Zheshen_15,barzanjeh2015microwave,chang2019quantum,zhang2020multidimensional,gregory2020imaging}, microscopy~\cite{Ono_2013}, biological sensing~\cite{taylor2013biological}, and phase tracking~\cite{yonezawa2012quantum} (see Refs.~\cite{lawrie2019quantum,pirandola2018advances,degen2017quantum,toth2014quantum,giovannetti2011advances} for a comprehensive review of quantum metrology). The focus of all these quantum-metrology experiments is on utilizing quantum resources to improve the performance at a single sensor. Recent theoretical works~\cite{ge2017distributed,proctor2017multi,zhuang2018distributed,eldredge2018optimal} proposed distributed quantum sensing (DQS) protocols that enable multiple sensors to leverage their shared entangled states to boost the performance of probing {\em global} properties of an interrogated object. Since a multitude of real-world applications rest upon a network or an array of sensors that work collectively to undertake sensing tasks, DQS significantly broadens the applicable scope of quantum metrology and opens a new route for achieving a quantum advantage with near-term quantum hardware.

This Review first provides in Sec.~\ref{sec:overview} an overview of DQS and compares it with distributed classical sensing (DCS).  Sec.~\ref{sec:protocols} describes DQS protocols with continuous-variables (CVs) and discrete variables (DVs). Sec.~\ref{sec:fisher} then analyzes DQS and DCS protocols, quantifies DQS's quantum advantage, and accounts for practical nonidealities. To articulate the applications of DQS, Sec.~\ref{sec:applications} reviews two recent DQS experiments for, respectively, optical phase sensing and radio-frequency (RF) sensing applications. An outlook for future prospects of DQS will be discussed in Sec.~\ref{sec:outlook}.

\section{Overview}
\label{sec:overview}

Usual multi-parameter estimation protocols endeavor to estimate {\em all} unknown parameters, while some multi-parameter estimation protocols alternatively consider extracting a single parameter while regarding the rest as nuisance parameters. DQS belongs to neither cases as it aims at estimating a global property of multiple parameters.

\begin{figure}[bth]
    \centering
    \includegraphics[width=1\textwidth]{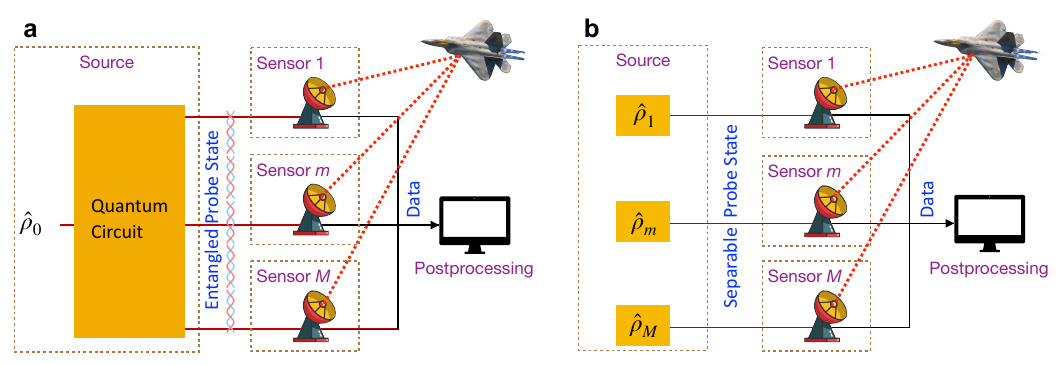}
    \caption{Distributed quantum sensing (a) with entangled inputs vs distributed classical sensing (b) with separable inputs. }
    \label{fig:DQS_concept}
\end{figure}

A conceptual schematic for DQS is sketched in Fig.~\ref{fig:DQS_concept}(a): a quantum circuit processes the initial quantum state $\hat{\rho}_0$ to create an entangled probe state shared by $M$ sensors that jointly interrogate an object. The measurement data produced by all sensors are postprocessed to infer a global parameter of the probed object. By contrast, in DCS illustrated in Fig.~\ref{fig:DQS_concept}(b), the quantum state at $M$ sensors is in a separable form (e.g., a product form $\hat{\rho}_1\otimes\hat{\rho}_2\otimes...\otimes\hat{\rho}_M$) to generate measurement data that are also postprocessed to infer a global property of the object. The ideal measurement sensitivity of DQS typically scales as $1/M$, whereas the scaling of the measurement sensitivity of DCS is limited by the SQL of $1/\sqrt{M}$.

A quantitative and equitable performance comparison between DQS and DCS typically relies on an appropriately chosen {\em resource counting} scheme. Resource counting is critical because DCS can otherwise always outperform DQS by, e.g., using more power at the sensors. Two resource-counting schemes have been widely employed. In the first scheme, the compared DQS and DCS protocols are assumed to use the same amount of total power at their sensors. The SQL is then defined as the measurement sensitivity of the DCS protocol subject to the power constraint. Such a resource-counting scheme has been adopted by other quantum-sensing protocols such as quantum illumination~\cite{Lopaeva_2013,Zheshen_15,barzanjeh2015microwave,chang2019quantum,zhang2020multidimensional,gregory2020imaging}. In the second resource-counting scheme, we first pick a DCS protocol operating with pure classical resources at a certain power level, based on which the SQL is derived. We then introduce nonclassical resources such as squeezed light and entanglement to transform the DCS protocol into a DQS protocol that overcomes the SQL. The second resource-counting scheme was used by LIGO~\cite{abadie2011gravitational,aasi2013enhanced,tse2019quantum} and by a recent DQS experiment~\cite{xia2020demonstration} (see Sec.~\ref{sec:applications}).






\section{Distributed Quantum Sensing Protocols}
\label{sec:protocols}

\begin{figure}[bth]
    \centering
    \includegraphics[width=0.5\textwidth]{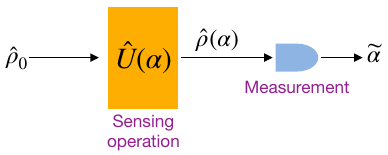}
    \caption{Sensing modeled as a single-parameter estimation problem. The initial state $\hat{\rho}_0$ is used to probe a parameter $\alpha$ embedded in the unitary operation $\hat{U}(\alpha)$, yielding output probe state $\hat{\rho}(\alpha)$. A measurement on $\hat{\rho}(\alpha)$ produces an estimator $\tilde{\alpha}$.}
    \label{fig:sensing_model}
\end{figure}

Prior to introducing DQS, let's first formulate quantum sensing using an example of single-parameter estimation, as shown in Fig.~\ref{fig:sensing_model}. Consider a parameter $\alpha$ embedded in a unitary operation $\hat{U}\left(\alpha\right)=\exp\left[-i \hat{H} \alpha\right]$, whose generator is the Hermitian operator $\hat{H}$. To sense $\alpha$, one prepares an initial quantum state $\hat{\rho}_0$ and passes it through $\hat{U}\left(\alpha\right)$, obtaining the output state 
\be
\hat{\rho}\left(\alpha\right) =\hat{U}\left(\alpha\right)\hat{\rho}_0\hat{U}^\dagger\left(\alpha\right)
\label{rho_alpha}
\ee 
after the sensing operation. Subsequently, a suitable measurement on $\hat{\rho}\left(\alpha\right)$ gives information for building an estimator $\tilde{\alpha}$. The precision of such an estimator is characterized by the variance, i.e., the mean squared error
$
\delta \alpha^2=\expval{\left(\tilde{\alpha}-\alpha\right)^2},
$ 
where $\delta \alpha$ is the standard deviation (std), i.e., the root-mean-square (rms) error.
To improve the precision of estimation, one can independently use $M$ probes to interrogate the same object and repeat the measurement process $M$ times. In this case, the initial state is in a product form of $\hat{\rho}_0^{\otimes M}$, and the measurement on each output state $\hat{\rho}\left(\alpha\right)$ yields the estimator $\tilde{\alpha}$. Due to the i.i.d. nature of individual sensing attempts, a simple average over all estimators leads to an overall precision of
$ 
\delta \alpha_M=\delta \alpha/\sqrt{M},
$ 
per the law of large numbers. Such a scaling is also called the SQL.

In a distributed sensing scenario, one looks at a more general problem: consider a general unitary operation comprising a product of $M$ unitary operations $\hat{U}\left(\bm \alpha\right)\equiv \otimes_{\ell=1}^M\hat{U}\left(\alpha_\ell\right)$, which, in general, entail $M$ unknown parameters. To carry out sensing, one inputs $M$ probes in a quantum state $\hat{\bp}_M$ and obtains the output quantum state
\be
\hat{\bp}_M\left(\bm \alpha\right)=\hat{U}\left(\bm \alpha\right)\hat{\bp}_M\hat{U}^\dagger\left(\bm \alpha\right),
\label{rho_output_general}
\ee
where the unknown parameters are $\bm \alpha=\left\{\alpha_m\right\}_{m=1}^M$ instead of being a single unknown parameter as in Eq.~\eqref{rho_alpha}. In a multi-parameter estimation scenario, one aims to estimate {\em all} unknown parameters; alternatively, one may only be interested in a single parameter while regarding the rest $M-1$ parameters nuisance. Distributed sensing belongs to neither categories, as it endeavors to estimate a global parameter, e.g., a weighted average
\be 
\bar{\alpha}=\bm w\cdot \bm \alpha\equiv \sum_{\ell=1}^M w_\ell \alpha_\ell,
\label{weighted_average}
\ee 
where the weights, $\bm w=\{w_m\}_{m=1}^M$, are non-negative and sum to one without loss of generality. One can in principle consider a general analytical scalar function $f\left(\bm \alpha\right)$, but as pointed out in Ref.~\cite{qian2019heisenberg}, via adaptive strategies one can obtain a rough estimate for $\tilde{\bm \alpha}$ and then perform linear expansion around the point to obtain $f\left(\bm \alpha\right)\simeq f\left(\tilde{\bm \alpha}\right)+\sum_\ell \partial_{\alpha_\ell} f\left(\tilde{\bm \alpha}\right) \left(\alpha_\ell-\tilde{\alpha}_\ell\right)$. By suitably dividing the number of measurements, one can reduce the problem of estimating the scalar function to estimating the weighted average in Eq.~\eqref{weighted_average}. 

We point out that the special case of identical parameters $\alpha_m=\alpha$ and identical weights $w_m=1/M$ has been considered in pioneering works of quantum metrology~\cite{Giovannetti_2001,giovannetti2004,giovannetti2006}, which show that entanglement can beat the SQL (See Sec.~\ref{sec:beat_SQL} for more details) and approach the Heisenberg limit of $\delta \alpha\propto 1/M$. A major advance made by DQS lies in the extension to a multi-parameter scenario, with an arbitrary scalar function of the parameters under estimation.

In the following, we will describe several DQS protocols and analyze their performances in the presence of imperfections.

\subsection{Distributed displacement sensing} 
\label{sec:distributed_displacement}
This Section covers single-quadrature displacement DQS protocols, with the generator being $\hat{H}=\hat{p}$. The protocol is based on entanglement generated from linear optics and single-mode squeezed vacuum states. Extension to a DQS protocol capable of simultaneous estimation of both quadrature displacements can be achieved by replacing the single-mode squeezed vacuum with a Gottesman–Kitaev–Preskill (GKP) grid state~\cite{gottesman2001encoding}, as studied in Ref.~\cite{zhuang2020distributed}.

\subsubsection{Special DQS protocols}
\label{CV_special}

\begin{figure}
\centering
\includegraphics[width=1\textwidth]{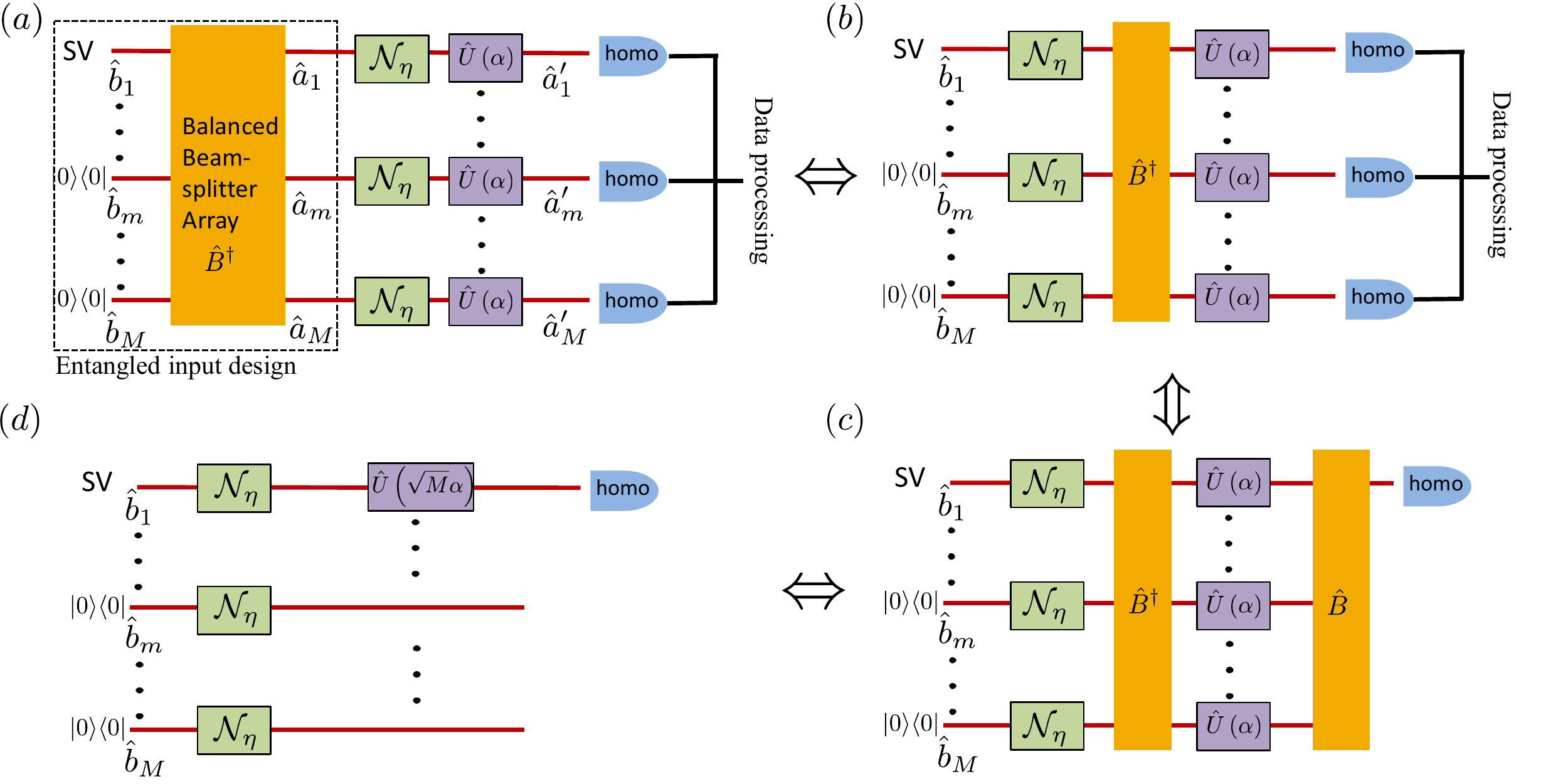}
\caption{Distributed displacement sensing with identical displacements and weights. SV: squeezed-vacuum state with mean photon number $N_S$ and squeezed real quadrature. $\mathcal{N}_\eta$:  pure-loss channel with transmissivity $0< \eta \le 1$.  $\hat{U}(\alpha)$:  field-quadrature displacement by real-valued $\alpha$.  homo: homodyne measurement on the real quadrature. (a) Original protocol. (b)--(d) Equivalent protocols from reduction relation (see main text for details).
\label{schematic_M}
}
\end{figure}

We begin with a special case of DQS with identical displacements and identical weights~\cite{zhuang2018distributed}. As shown in Fig.~\ref{schematic_M}(a), the entangled probe state is prepared by distributing a single-mode squeezed vacuum state $\hat{b}_1$ on a balanced beam-splitter array. After each probe mode $\hat{a}_m$ going through the displacement $\hat{U}\left(\alpha\right)$, one performs a homodyne measurement on the real quadrature ${\rm Re}\left(\hat{a}_m^\prime\right)$. Combining all measurement 
data $\{\tilde{a}_m^\prime\}_{m=1}^M$ through classical data processing, one obtains an estimator $\tilde{\alpha}=\sum_{m=1}^M \tilde{a}_m^\prime/M$. We model the imperfection in each sensor by the most common model of a pure-loss channels $\calN_\eta$ of transmissivity $\eta$, which is described by the Bogoliubov transform
\be 
\hat{a}\to \sqrt{\eta}\hat{a}+\sqrt{1-\eta} \hat{e},
\ee 
on the input mode $\hat{a}$ and vacuum environment mode $\hat{e}$. To begin with, we consider the losses to be identical across different sensors.

To analyze the performance, a sequence of reduction is utilized to simplify the problem to a single-parameter estimation situation. First, since the beam-splitter array $\hat{B}^\dagger$ commutes with the identical loss channels $\calN_\eta^{\otimes M}$, an equivalent protocol is devised in Fig.~\ref{schematic_M}(b). Moreover, the data processing of equal weighted sum can be implemented by a balanced beam-splitter array $\hat{B}$ in front of the homodyne measurements, leading to another equivalent protocol in Fig.~\ref{schematic_M}(c), where the identical displacements are now sandwitched between two beam-splitter arrays such that a single homodyne measurement suffices to carry out the estimation. At this juncture, we can utilize a unitary reduction relation
\be 
\hat{B}^\dagger \hat{U}^{\otimes M}\left(\alpha\right) \hat{B}=\hat{U}\left(\sqrt{M}\alpha\right) \otimes \hat{U}^{\otimes \left(M-1\right)}\left(0\right)
\ee 
to further simplify the protocol to Fig.~\ref{schematic_M}(d), which shows that the multimode protocol is reduced to a single-mode protocol, as only $\hat{b}_1$ interrogates a single channel that entails information about $\alpha$. A single-mode displacement is estimated after the reduction, thus results in Sec.~\ref{opt_Gaussian} immediately infer that the optimal Gaussian~\cite{Weedbrook2012} input state for the reduced protocol depicted in Fig.~\ref{schematic_M}(d) is a single-mode squeezed vacuum state. The resulting sensitivity obeys
\be 
\delta \alpha_\eta^{\rm E}=\frac{1}{\sqrt{I_F^{\rm Gauss, E, max}}}=\left(\frac{\eta}{M\left(\sqrt{N_S+1}+\sqrt{N_S}\right)^2}+\frac{1-\eta}{M}\right)^{1/2},
\label{precision_lossy_E_simple}
\ee 
where $I_F^{\rm Gauss, E, max}$ is given in Eq.~\eqref{F_var_u_CV_optimal_loss}. We note that although the input state in the reduced protocol is a single-mode squeezed-vacuum state, the $M$-mode input $\{\hat{a}_m\}_{m=1}^M$ in the original DQS protocol illustrated Fig.~\ref{schematic_M}(a) is in a multipartite entangled state. The sensitivity given by Eq.~\eqref{precision_lossy_E_simple} is the optimum among all DQS protocols based on {\em Gaussian} states. Moreover, in a lossless situation ($\eta=1$), the above DQS protocol is the optimum among {\em all} DQS protocols per the results introduced in Sec.~\ref{sec:CV_Fisher}.

\begin{figure}
\centering
\subfigure{
\includegraphics[width=0.36\textwidth]{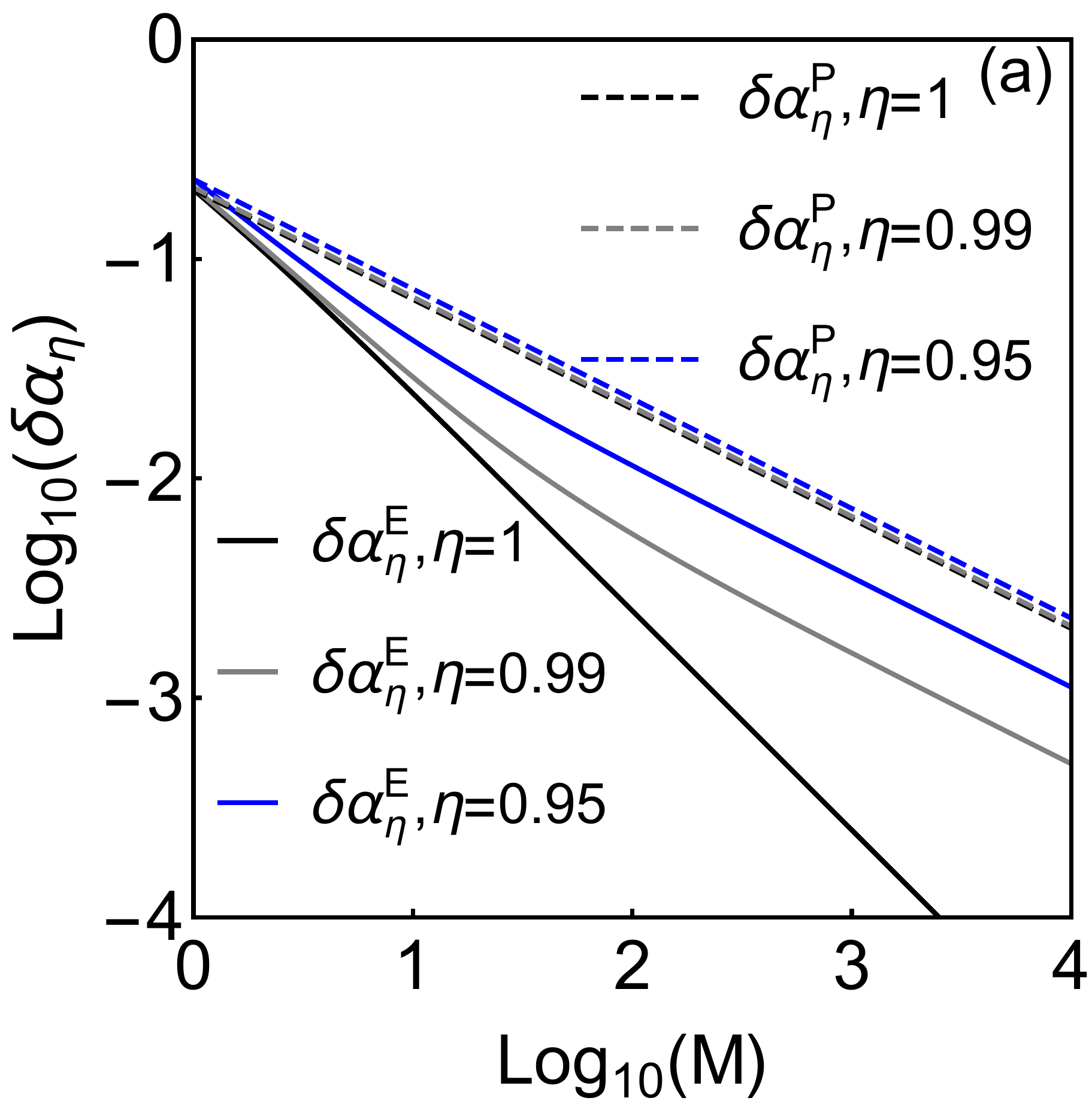}
\label{Heisenberg}
}
\subfigure{
\includegraphics[width=0.38\textwidth]{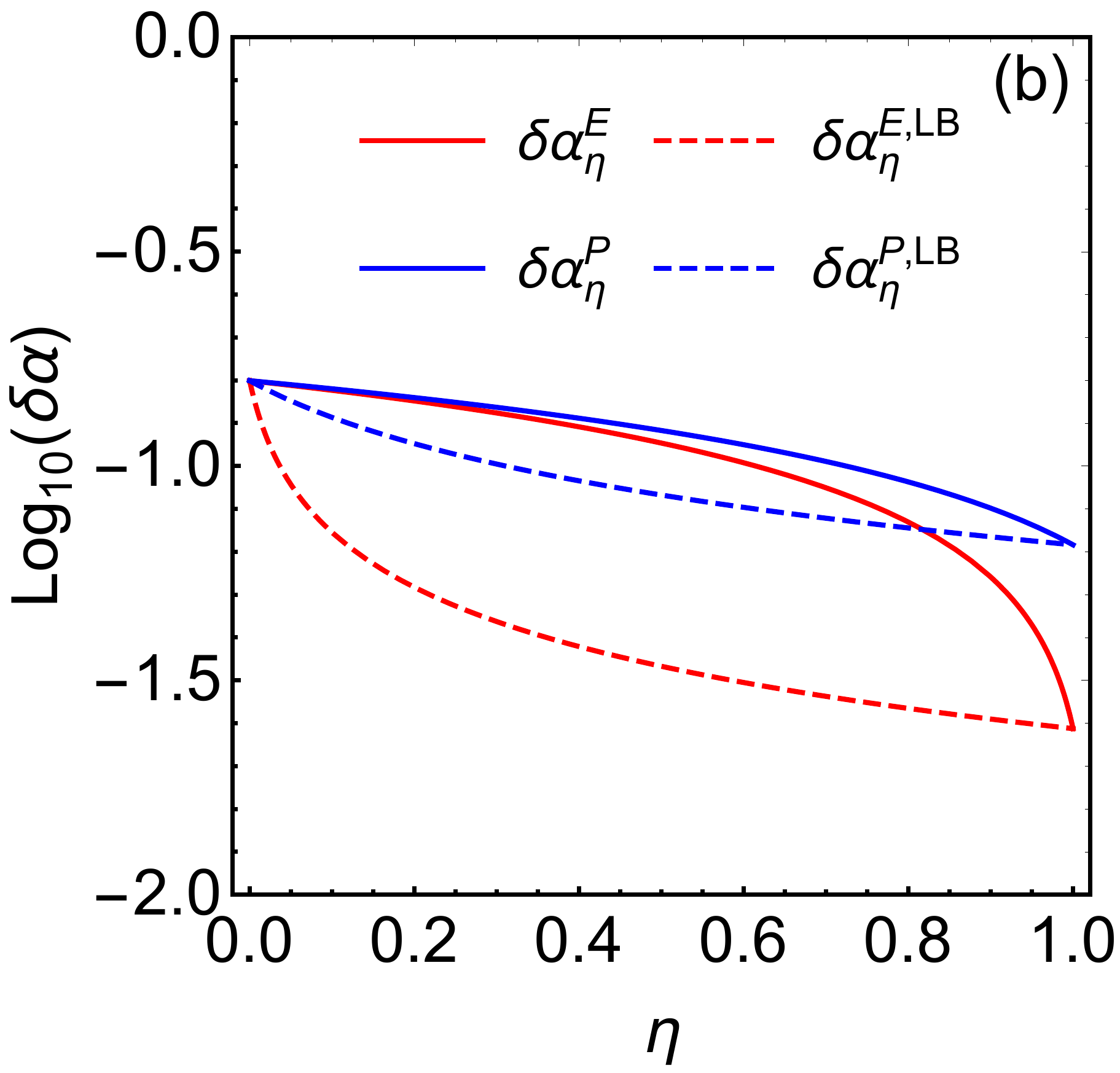}
\label{bound_compare}
}
\caption{(a) Plots of the rms estimation errors (Reproduced from Ref.~\cite{zhuang2018distributed}), $\delta\alpha_\eta^E$ (solid curves) and $\delta\alpha_\eta^P$ (dashed curves), versus the number of sensor nodes, $M$, for various transmissivity values, from top to bottom, $\eta = 0.95, 0.99, 1$, with $n_S \equiv N_S/M$ fixed at $n_s = 1$. 
(b) Comparison between bounds and the performances (Reproduced from Ref.~\cite{xia2018repeater}). $N_S=10, M=10$, $\delta \alpha_\eta^{\rm E}$: performance of the optimal entangled Gaussian scheme for DQS (red). $\delta \alpha_\eta^{\rm P}$: performance of optimal separable Gaussian scheme for DCS (blue). $\delta \alpha_\eta^{\rm E, LB}$: lower bound of precision in Eqs.~\eqref{E_bound} for all entangled scheme (red, dashed). $\delta \alpha_\eta^{\rm P,LB}$: lower bound of precision in Eq.~\eqref{C_bound} for all separable scheme. When $\eta\sim 1$, we see that $\delta \alpha_\eta^{\rm E}$ beats the non-tight lower bound $\delta \alpha_\eta^{\rm P,LB}$ of separable scheme.
\label{new bound}
}
\end{figure}

Similarly, subject to the resource-counting of mean photon number, the optimal DCS protocol with Gaussian states for $\eta<1$ and the general optimal DCS protocol for $\eta=1$ is based on a product of single-mode squeezed-vacuum states and yields a sensitivity level of
\be 
\delta \alpha_\eta^{\rm P}=\frac{1}{\sqrt{I_F^{\rm Gauss, C, max}}}
= \left(\frac{\eta}{M\left(\sqrt{N_S/M+1}+\sqrt{N_S/M}\right)^2}+\frac{1-\eta}{M}\right)^{1/2},
\label{precision_lossy_P_simple}
\ee 
where $I_F^{\rm Gauss, C, max}$ is given by Eq.~\eqref{F_varU_product_CV_loss}. The performance of the optimal DQS protocol and the optimal DCS protocol is compared in Fig.~\ref{Heisenberg}. At $\eta=1$ and a fixed $n_S\equiv N_S/M$, Eqs.~\eqref{precision_lossy_E_simple} and~\eqref{precision_lossy_P_simple} indicate that $\delta \alpha_\eta^{\rm E}\sim 1/M$, i.e., a Hisenberg scaling, while $\delta \alpha_\eta^{\rm P}\sim 1/\sqrt{M}$ is subject to the SQL, as is verified in Fig.~\ref{Heisenberg}. For sub-unity $\eta$, $\delta \alpha_\eta^{\rm E}$ deviates from the Heisenberg scaling at large $M\propto 1/(1-\eta)$. To sustain the Heisenberg scaling at a large $M$, one can utilize bosonic error-correcting codes, as demonstrated in Ref.~\cite{zhuang2020distributed}.

Now that we know the optimality of the performances in Eqs.~\eqref{precision_lossy_E_simple} and~\eqref{precision_lossy_P_simple} in special cases, We can also  compare them with the general precision lower bounds
\ba
\delta \alpha_\eta^{\rm E,LB}=\frac{1}{\sqrt{I_F^{\rm UB,E}}}=\left(\eta 4M \left(\sqrt{N_S}+\sqrt{N_S+1}\right)^2+4\left(1-\eta\right)M\right)^{-1/2}
\label{E_bound}
\\
\delta \alpha_\eta^{\rm C,LB}=\frac{1}{\sqrt{I_F^{\rm UB,C}}}=\left(\eta 4M \left(\sqrt{N_S/M}+\sqrt{N_S/M+1}\right)^2+4\left(1-\eta\right)M\right)^{-1/2},
\label{C_bound}
\ea
which are obtained from the Fisher information upper bounds Eqs.~\eqref{F_var_u_CV_optimal_UB_loss} and~\eqref{F_varU_product_CV_UB_loss} in Sec.~\ref{general_bound}. We compare the above bounds and the actual protocol's performance in Fig.~\ref{bound_compare}. Indeed, the lower bounds are tight at $\eta=1$ (also trivially tight at $\eta=0$), confirming the optimality. However, when $\eta<1$, there is a gap between the optimal Gaussian protocol and the lower bound. It is an open question whether other protocols can beat the optimal Gaussian protocol.

\subsubsection{General DQS protocols}
\label{sec:CV_distributed_sensing}
\begin{figure}
\centering
\includegraphics[width=0.5\textwidth]{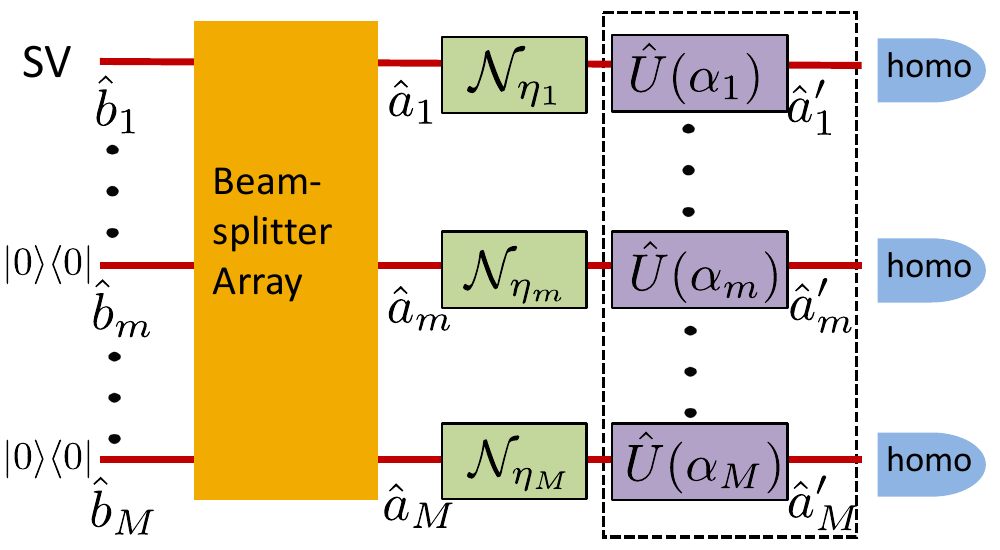}
\caption{Distributed displacement sensing in the general case. $\mathcal{N}_{\eta_m}$:  pure-loss channel with transmissivity $0< \eta_m \le 1$.  $\hat{U}(\alpha_m)$:  field-quadrature displacement by real-valued $\alpha_m$.  
\label{int1}
}
\end{figure}
We now extend the special DQS protocol of Sec.~\ref{CV_special} to a DQS protocol for estimating an arbitrary weighted average of heterogeneous displacements in the presence of heterogeneous experimental imperfections. As shown in Fig.~\ref{int1}, the objective is to estimate a weighted average, $\bar{\alpha}\equiv \sum_{m=1}^M w_m \alpha_m,$ of displacements $\{\alpha_m\}_{m=1}^M$ among $M$ sensor nodes. Experimental imperfections are modelled by lossy channels $\{\calN_{\eta_m}\}_{m=1}^M$, where the transmissivities $\bn = \{\eta_m\}_{m=1}^M$ can vary across different sensor nodes. The input modes $\{\hat{a}_m\}_{m=1}^M$ are in an entangled state, generated from distributing a single-mode squeezed-vacuum state via a general beam-splitter array with ratios suitably optimized.

\begin{figure}
\centering
\includegraphics[width=1\textwidth]{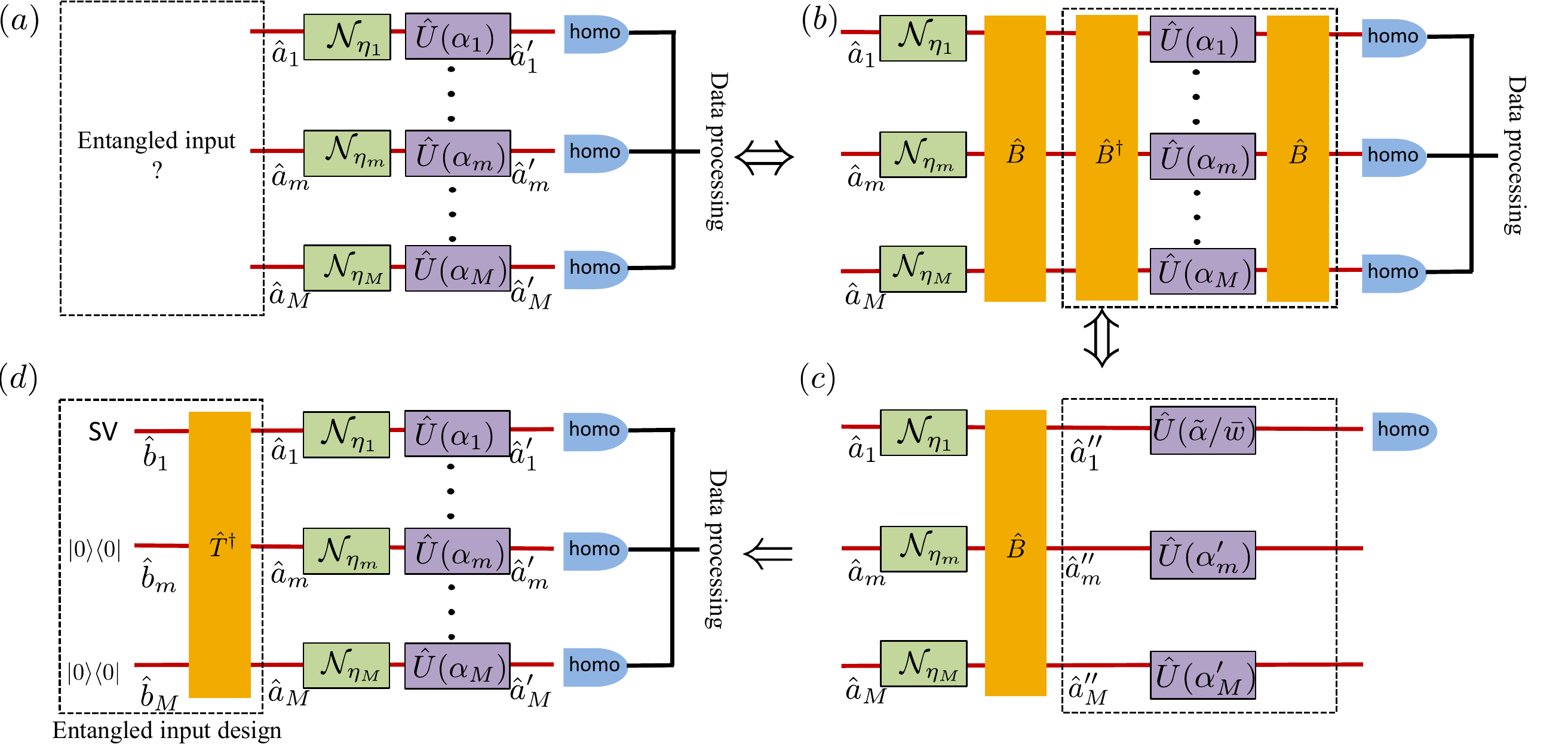}
\caption{Schematic of the analyses. (a) Original protocol with unknown optimal entangled input. (b)-(d) Equivalent protocols from reduction relation (see main text for details). 
\label{analyses}
}
\end{figure}

To derive the optimal entangled probe state, we again utilize equivalence relations shown in Fig.~\ref{analyses}. Fig.~\ref{analyses}(a) illustrates the problem of deriving the optimal entangled probe state assuming homodyne measurements. As shown in Fig.~\ref{analyses}(b), before the homodyne measurements, one can apply a beam-splitter array $\hat{B}$, which can be compensated in the data processing on measurement data while preserving the Fisher information. Moreover, one can insert $\hat{B}\hat{B}^\dagger=\hat{I}$ in between the lossy channels and the sensing displacement operations. From properties of Gaussian unitary, as shown in Fig.~\ref{analyses}(c) one can suitably choose $\hat{B}$ to obtain
\be 
\hat{B}^\dagger \left[\otimes_{m=1}^M \hat{U}\left(\alpha_m\right) \right]\hat{B}=\hat{U}\left(\bar{\alpha}/\bar{w}\right)\otimes \left[\otimes_{m=2}^M \hat{U}\left(\alpha_m^\prime\right) \right],
\ee 
where $\bar{w} \equiv \sqrt{\sum_{m=1}^Mw_m^2}$ and all other parameters $\alpha_m^\prime$ are independent of the parameter of interest $\bar{\alpha}$. As such, one only needs to measure the first output mode to estimate $\bar{\alpha}$. Now the problem becomes choosing the optimal input among $\{\hat{a}_m\}_{m=1}^M$ such that mode $\hat{a}_1^{\prime \prime}$ is the optimum for single-mode displacement sensing. Considering the transform $\hat{B}$ and the lossy channels, it is easy to obtain
\be 
\hat{a}_1^{\prime \prime}=\sum_{m=1}^M \frac{w_m}{\bar{w}}\sqrt{\eta_m}\hat{a}_m+\mbox{ vacuum terms}=\sqrt{\bar{\eta}}\hat{b}_1+\mbox{ vacuum terms},
\ee 
where $\hat{b}_1 \equiv \sum_{m=1}^Mw_m\sqrt{\eta_m}\,\hat{a}_m/\bar{w}$,  $\bar{\eta} \equiv \sum_{m=1}^M w_m^2\eta_m/\bar{w}^2$, and $\bar{w} \equiv \sqrt{\sum_{m=1}^M  w_m^2\eta_m}$. This indicates that the effective probe mode $\hat{a}_m^{\prime \prime}$ is obtained from passing a single mode $\hat{b}_1$ through a pure-loss channel $\calN_{\bar{\eta}}$. Given the energy constraint, the optimal Gaussian state is therefore obtained by applying a beam-splitter array $\hat{T}^\dagger$ on a single-mode squeezed vacuum mode $\hat{b}_1$ and vacuum states, as shown in Fig.~\ref{analyses}(d).

With the above protocol, $\tilde{\alpha} \equiv \sum_{m=1}^Mw_m{\rm Re}(\hat{a}'_m)$ is an unbiased estimator of $\bar{\alpha}$ with the minimum rms error
\be
\delta\alpha_{\bm \eta}^{\rm E} = \frac{\bar{w}}{2}\left(\frac{\bar{\eta}}{(\sqrt{N_S+1}+\sqrt{N_S})^2}+1-\bar{\eta}\right)^{1/2}.
\label{dalpha_extension}
\ee
Parallel to Sec.~\ref{CV_special}, one can compare the above DQS protocol with the optimal Gaussian DCS protocol, which, under the considered scenario, employs a product of single-mode squeezed-vacuum states, yielding a precision of
\be
\delta \alpha_{\bm \eta}^{\rm P}=\min_{\sum_{m=1}^M N_m=N_S }
\left[\sum_{m=1}^Mw_m^2\left(\frac{\eta_m}{\left(\sqrt{N_m+1}+\sqrt{N_m}\right)^2}+1-\eta_m\right)/4\right]^{1/2}
\label{dalpha_product_extension}
\ee
without a known closed form. One can easily check that at $\eta_m=\eta$, $\alpha_m=\alpha$, $w_m=1/\sqrt{M}$, Eqs.~\eqref{dalpha_extension} and~\eqref{dalpha_product_extension} converge to the simple case of Eqs.~\eqref{precision_lossy_E_simple} and~\eqref{precision_lossy_P_simple}. Also, we emphasize that the same optimality holds: the above DCS protocol is the optimum among all protocols based on {\em Gaussian} states and the optimum among {\em all} DCS protocols at $\eta_m=1$ for all $m$.

\subsection{Distributed phase sensing}
\label{variation:phase}

In distributed phase sensing, Eq.~\eqref{rho_output_general} involves a phase rotation unitary operation $\hat{U}\left(\alpha_m\right)=\exp\left(-i\alpha_m \hat{a}_m^\dagger \hat{a}_m\right)$ at each probe $\hat{a}_m$. As explained in Sec.~\ref{sec:CV_Fisher}, the Fisher information of phase sensing is unbounded for non-Gaussian states if only subject to a photon-number constraint. As such, previous works have been focusing on finding the optimal Gaussian protocol~\cite{oh2019optimal,oh2020optimal}, considering constraints on the photon-number variance~\cite{escher2011,gagatsos2017bounding} or limiting to a finite photon-number cutoff~\cite{dorner2009optimal,demkowicz2009quantum}.

Through evaluating the precision limit of DQS in Eq.~\eqref{CR_distributed} for all Gaussian states (obtained from the multi-parameter CR bound, see Sec.~\ref{sec:CR_multi}), Ref.~\cite{oh2020optimal} shows that the CV entangled state from distributing a single-mode squeezed-vacuum state via a beam-splitter array---identical to the optimal state in the displacement case in Sec.~\ref{sec:distributed_displacement}---is the optimal Gaussian state for estimating the weighted sum of phase. 

In the identical weight case, a balanced beam-splitter array is adopted and the corresponding precision
\be 
\delta \alpha^E=\frac{1}{\sqrt{8N_S\left(N_S+1\right)}}=\frac{1}{\sqrt{8M n_S\left(M n_S+1\right)}},
\label{precision_phase_E}
\ee 
can be easily achieved by performing homodyne measurements---again identical to the approach in the displacement DQS of Sec.~\ref{sec:distributed_displacement}. While the optimal separable input and homodyne measurement (identical to the case of  Sec.~\ref{sec:distributed_displacement}) gives the performance
\be 
\delta \alpha^P=\frac{1}{\sqrt{8 M n_S\left(n_S+1\right)}}.
\ee 
We see a similar advantage from the entanglement as the case of displacement DQS discussed in Sec.~\ref{sec:distributed_displacement}. Although the above state is the optimal for an ideal lossless situation, in the presence of loss, however, Ref.~\cite{guo2020distributed} finds that adding a non-zero displacement to the original single-mode squeezed vacuum can further improve the performance. The optimal entangled probe state in the presence of loss remains unknown, even restricting to Gaussian states.

Like the displacement DQS protocols, the resource-counting scheme should be properly chosen to ensure an equitable comparison with DCS protocols. In this regard, the above works have adopted the resource-counting scheme of the total mean photon number, accounting for the coherent-state portion and the squeezed-state portion, i.e., the first resource-counting scheme introduced in Sec.~\ref{sec:overview}. This resource-counting scheme is justified if the power probing the samples has to be limited, for example, for the purpose of covertness or avoiding harming delicate biological tissues. However, coherent states are much easier to generate than squeezed-vacuum states and is often orders of magnitude higher in power as well. In applications, e.g., LIGO, that require strong coherent light, resource counting is on the power of the coherent-state portion, i.e., the second resource-counting scheme discussed in Sec.~\ref{sec:overview}. In some phase DQS protocols, phase shifts transform strong coherent states into displacements so that the displacement DQS protocol introduced in Sec.~\ref{sec:distributed_displacement}) can then be utilized, as proposed in Ref.~\cite{zhuang2018distributed}. A similar technique is adopted in DQS for RF-sensing applications, as detailed in Sec.~\ref{sec:rf}.

As we see in the above protocols for phase sensing and displacement sensing, a beam-splitter array, i.e., a linear optical network, has the power of generating multipartite entangled states that benefit estimating global parameters. In this regard, Ref.~\cite{ge2017distributed} systematically evaluated the power of entangled states prepared by a beam-splitter array on separable inputs $\otimes \ket{\psi_m}_{m=1}$.

\begin{figure}
\centering
\includegraphics[width=0.5\textwidth]{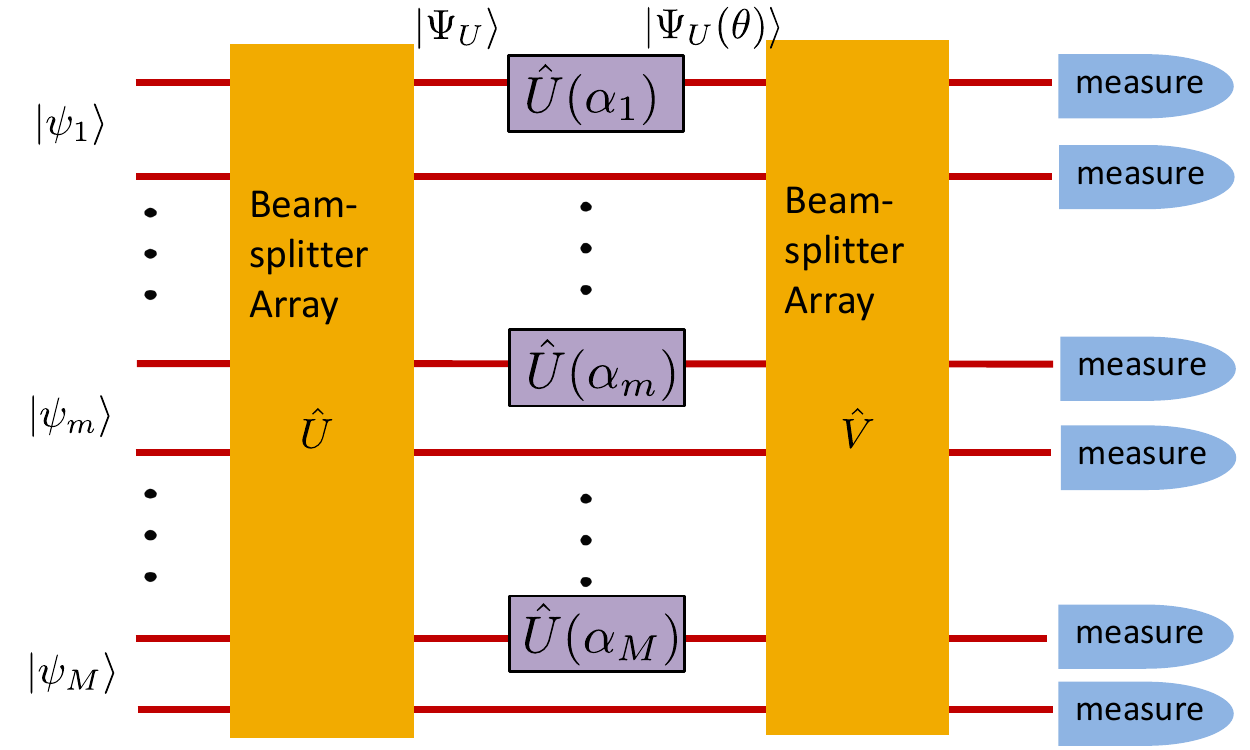}
\caption{General distributed phase sensing with beam-splitter array and product inputs (see also Fig. 1(b) in Ref.~\cite{ge2017distributed}). 
\label{fig:ge}
}
\end{figure}

A different resource-counting scheme needs to be adopted for phase sensing because the precision can be unbounded under a simple mean photon-number constraint. A commonly used resource-counting scheme for phase sensing is based on the photon-number fluctuation,
\be 
n_j^{\rm LO}\equiv \sqrt{\ev{\hat{n_j}^2}{\psi_j}}\le n_j^\star,
\ee 
of each two-mode product state $\ket{\psi_j}$ as an input to the beam-splitter array. For general {\em unnormalized} weights $\bm w^\star$ ($\max_j |w_j^\star|=1/d$, which is different from our previous definition) and photon-number fluctuation constraints $\bm n^\star\equiv \left(n_1^\star,\cdots, n_M^\star\right)$, Ref.~\cite{ge2017distributed} showed that the precision of phase DQS is lower bounded by
\be 
\delta \bar{\alpha}\ge \delta \bar{\alpha}^{\rm LB*}\equiv \frac{M |\bm w^\star|^2}{2|\bm n^\star|},
\ee 
where $|\bm w^\star|^2=\sum_j w_j^{\star2}$ and $|\bm n|^2=\sum_j n_j^{\star2}$. To better understand the interpretation, we can consider the equal weight case, i.e., $|\bm w^\star|^2=1/M$, yielding
\be 
\delta \bar{\alpha}\ge \delta \bar{\alpha}^{\rm LB*}=\frac{1}{2|\bm n^\star|}\ge \frac{1}{2N_S^\star}.
\label{bound_Ge}
\ee 
The second lower bound is achieved by concentrating all $N_S^\star$ photons on the same mode, so $|\bm n^\star|=N_S^\star=M n_S^\star$ saturates the Heisenberg limit. While an even distribution of photons among all input modes of the beam-splitter array leads to the SQL $\delta \bar{\alpha}\ge1/(2n_S^\star\sqrt{M})$. As such, the benefit from entanglement only arises when photons are concentrated on a few input modes to the beam-splitter array. A protocol is also provided in Ref.~\cite{ge2017distributed} by choosing the input to the beam-splitter array as two photon-number state.

Now let's further compare the lower bound in Eq.~\eqref{bound_Ge} with the achievable performance in Eq.~\eqref{precision_phase_E}. Since $N_S^\star=\sqrt{\expval{\hat{n}^2}}=\sqrt{N_S\left(3N_S+2\right)}$ for a single-mode squeezed-vacuum state, Eq.~\eqref{bound_Ge} yields a valid lower bound 
\be 
\delta \bar{\alpha}^{\rm LB*}=\frac{1}{2 \sqrt{N_S\left(3N_S+2\right)}} \le \delta \alpha^E=\frac{1}{\sqrt{8N_S\left(N_S+1\right)}}.
\ee 
In the $N_S\gg1$ limit,
\be 
\frac{\delta \alpha^E}{\delta \bar{\alpha}^{\rm LB*}}\sim \sqrt{\frac{3}{2}}.
\ee 
Note that the DV DQS protocol based on generalized twin-Fock state~\cite{ge2017distributed} gives $\delta \alpha^E=2/\sqrt{2N_S(N_S+2)}$. Although also achieving a Heisenberg scaling, the DV DQS protocol's performance is inferior to that of the CV DQS protocols~\cite{oh2020optimal,zhuang2018distributed}.

\section{Performance Limits}
\label{sec:fisher}

In this Section, we analyze the performance limits for DQS protocols.
A brief review on the limits of quantum sensing, including quantum Cram\'{e}r-Rao (CR) bound, standard quantum limit (SQL) for separable protocols, and Heisenberg limit of entangled protocols will first be provided. Then, the quantum CR bound will be utilized to understand the performance limits of the DQS and DCS protocols in Sec.~\ref{sec:protocols}, which leads to conclusions about the optimality of these protocols.

\subsection{Quantum Cram\'{e}r-Rao bound and the standard quantum limit}
\label{sec:single_parameter_CR}

The ultimate precision of estimating a single-parameter $\alpha$ (without bias) from the state $\hat{\rho}\left(\alpha\right)$ is asymptotically given by the quantum CR bound~\cite{Helstrom_1976,Holevo_1982,Yuen_1973,paris2009quantum}
\be 
\delta \alpha_M^2 \ge  \frac{1}{M} \delta \alpha^2_{CR} \equiv \frac{1}{M I_F[\hat{\rho}\left(\alpha\right)]},
\label{CR_bound}
\ee 
where $M$ is the number of sensing attempts and the single-parameter Fisher information
\be
I_F[\hat{\rho}(\alpha)]\equiv  \lim_{\epsilon\to 0} 8\!\left\{1-
\sqrt{{\mathcal F}[\hat{\rho}(\alpha),\hat{\rho}(\alpha+\epsilon)]}\right\}/{\epsilon^2}
\label{F_Uhlmann}
\ee
is given in terms of the 
Uhlmann fidelity,  ${\mathcal F}(\hat{\sigma}_1,\hat{\sigma}_2)\equiv \left[{\rm Tr}\!\left(\sqrt{\sqrt{\hat{\sigma}_1}\hat{\sigma}_2\sqrt{\hat{\sigma}_1}}\right)\right]^2$, between states $\hat{\sigma}_1$ and $\hat{\sigma}_2$. Alternatively, the Fisher information can be obtained from introducing the symmetric logarithmic derivative (SLD) $\hat{L}_\alpha$, which is Hermitian and satisfies
\be 
\frac{\hat{L}_\alpha \hat{\rho}_\alpha+ \hat{\rho}_\alpha \hat{L}_\alpha}{2}=\frac{\partial \hat{\rho}_\alpha}{\partial \alpha}.
\label{LSD}
\ee 
Then in general the Fisher information is defined as
\be
I_F[\hat{\rho}(\alpha)]\equiv  {\rm Tr}\left[\hat{\rho}_\alpha \hat{L}_\alpha^2\right].
\label{F_LSD}
\ee
In fact, for quantum states in the form of Eq.~\eqref{rho_alpha}, one can easily obtain the Fisher information in both ways for pure states. The fidelity for a pure state $\hat{\rho}=\ketbra{\psi}$ is easily devried by
${\mathcal F}(\hat{U}\left(\alpha\right)\ket{\psi},\hat{U}\left(\alpha+\epsilon\right)\ket{\psi})
= |\ev{\hat{U}\left(\epsilon\right)}{\psi}|^2
$, and thus by taking derivatives the Fisher information reads
\be
I_F[\hat{U}\left(\alpha\right)\ket{\psi}]/4
={\rm var} \left(\hat{H} \right)_{\psi}\equiv \ev{\hat{H}^2}{\psi}-\ev{\hat{H}}{\psi}^2.
\label{F_var}
\ee 
In terms of SLD, a pure state has $\hat{\rho}_\alpha^2=\hat{\rho}_\alpha$, it is thus easy to verify that $\hat{L}_\alpha=2\partial_\alpha \hat{\rho}_\alpha=-2i \hat{H}\hat{\rho}_\alpha+2i \hat{\rho}_\alpha\hat{H}$, and therefore Eq.~\eqref{F_LSD} leads to Eq.~\eqref{F_var}. However, it is important to note that Eq.~\eqref{F_var} does not in general hold for mixed states, while both Eqs.~\eqref{F_Uhlmann} and~\eqref{F_LSD} remain valid for all mixed states. Moreover, the CR bound in Eq.~\eqref{CR_bound} is guaranteed to be asymptotically, i.e., at $M\gg1$, achievable for a single-parameter estimation problem by adaptive strategies~\cite{barndorff2000fisher,gill2005state,hayashi2006quantum,fujiwara2011strong,zhou2020saturating}, in which the input state and the associated measurement setting evolve as the number of measurement increases. However, all input states are constrained to a product form.

One can generalize the above protocol to using an arbitrary separable input state across $M$ measurements and allowing an arbitrary joint measurement~\cite{giovannetti2006}, however, the CR bound in Eq.~\eqref{CR_bound} holds true, and the $\delta \alpha_M\propto 1/\sqrt{M}$ scaling is usually referred to as the SQL. 

\subsection{Beating the standard quantum limit}
\label{sec:beat_SQL}

Entanglement between the input states across different measurements is needed to beat the SQL. Early works of quantum metrology~\cite{Giovannetti_2001,giovannetti2004,giovannetti2006} found that the so-called Heisenberg scaling of precision, $\delta \alpha_M \propto 1/M$, can be approached if the input states across $M$ measurements are in a multipartite entangled form. To analyze this setting, we consider applying a joint operation on a $M$-probe pure state $\ket{\bm \psi_M}$:
\be
\hat{\bp}_M\left(\alpha\right)=\hat{U}^{\otimes M}\left(\alpha\right)\ketbra{\bm \psi_M}\hat{U}^{\dagger\otimes M}\left(\alpha\right),
\label{rho_output}
\ee 
where $\hat{U}^{\otimes M}\left(\alpha\right)=\exp\left[-i \alpha \sum_\ell \hat{H}_\ell \right]$, with $\{\hat{H}_\ell\}$ being identical generators applied on different probes.
Then, the Fisher information is derived using Eq.~\eqref{F_var} as
\be
I_F[\hat{U}^{\otimes M}\left(\alpha\right)\ket{\bm \psi_M}]/4
= {\rm var} \left(\sum_\ell \hat{H}_\ell \right)_{\bm \psi_M}
\equiv \ev{(\sum_\ell \hat{H}_\ell)^2}{\bm \psi_M}-\ev{\sum_\ell \hat{H}_\ell}{\bm \psi_M}^2.
\label{F_var_u}
\ee 
For a product state $\ket{\bm \psi_M}=\otimes_{\ell=1}^M \ket{\psi_\ell}$ (not necessarily identical), $ \ev{(\sum_\ell \hat{H}_\ell)^2}{\bm \psi_M}= \sum_\ell \ev{ \hat{H}_\ell^2}{\psi_\ell}+\sum_{\ell \neq \ell^\prime}\ev{ \hat{H}_\ell }{\psi_\ell} \ev{\hat{H}_{\ell^\prime}}{\psi_{\ell^\prime}}$, which, after some simple algebra, shows additivity for Fisher information between product states:
\begin{align}
I_F[\hat{U}^{\otimes M}\left(\alpha\right)\otimes_{\ell=1}^M \ket{\psi_\ell}]/4
&=\sum_\ell  I_F[\hat{U}\left(\alpha\right) \ket{\psi_\ell}]/4.
\label{F_varU_product}
\end{align} 
Combining the convexity property of the Fisher information, we conclude that any separable input state will have Fisher information upper bounded by $M$ times the maximum Fisher information for the individual states, therefore giving the SQL of $\delta \alpha_M\propto 1/\sqrt{M}$. Eq.~\eqref{F_varU_product}, however, does not hold for entangled states. The disparity between the SQL performance and sub-SQL performance demonstrates a provable quantum advantage enabled by entanglement, as elaborated below by a few examples.

\subsubsection{Finite-dimensional probes}

For probe states of a finite $d$ dimensions, e.g., DV systems such as atoms, NV centers etc., or CV systems with a finite photon-number cutoff, one can simply consider the eigenstates $\{\ket{\lambda_n}\}_{n=1}^d$ of the generator, i.e., $\hat{H}\ket{\lambda_n}=\lambda_n \ket{\lambda_n}$, as discussed in Ref.~\cite{giovannetti2004}. Suppose the minimum and maximum eigenvalues are $\lambda_{\rm min}$ and $\lambda_{\rm max}$, each term $I_F[\hat{U}\left(\alpha\right) \ket{\psi_\ell}]/4={\rm var} \left(\hat{H} \right)_{\psi_\ell}\le \left(\lambda_{\rm max}-\lambda_{\rm min}\right)^2/4$. Eq.~\eqref{F_varU_product} then provides the following bound for a separable input
\be 
I_F\le I_F^{\rm max,C}\equiv M \left(\lambda_{\rm max}-\lambda_{\rm min}\right)^2.
\label{F_varU_product_DV}
\ee 
In contrast, with an entangled input, $\hat{H}_M=\sum_\ell \hat{H}_\ell$ is regarded as a global generator, and the minimum and maximum eigenvalues then become $M\lambda_{\rm min}$ and $M\lambda_{\rm max}$, leading to the Fisher information
\be 
I_F\le I_F^{\rm max,E}\equiv M^2\left(\lambda_{\rm max}-\lambda_{\rm min}\right)^2,
\label{F_varU_HS_DV}
\ee 
achievable by the superposition state $\ket{\bm \psi_M}=\left(\ket{\lambda_{\rm max}}^{\otimes M}+\ket{\lambda_{\rm min}}^{\otimes M}\right)/\sqrt{2}$. The above Fisher information follows $\propto M^2$, corresponding to an optimal Heisenberg scaling of $\delta \alpha_M\propto 1/M$~\cite{zwierz2010general}.

\subsubsection{Infinite-dimensional probes}
\label{sec:CV_Fisher}

For infinite-dimensional probe states, e.g., in a CV system, the eigenvalues of the generator can be unbounded. In this case, additional constraints, such as the mean occupation number (the mean photon number in the context of quantum optics) $N_S=\sum_{\ell=1}^M \expval{\hat{a}_\ell^\dagger\hat{a}_\ell}$, are necessary to ensure a physical input state.

The Fisher information for CV states needs to be calculated on a case-by-case basis. For example, in single-mode phase sensing with $\hat{H}=\hat{a}^\dagger \hat{a}$ being the number operator, one can construct an input state $\ket{\psi}=\sqrt{1-{N_S}/{S}}\ket{0}+\sqrt{{N_S}/{S}}\ket{S}$, with mean photon number $N_S$ bounded but the variance ${\rm var} \left(\hat{H} \right)_{\psi}=N_S (S-N_S)$ can be arbitrarily large, and therefore, the Fisher information can be unbounded. Hence, further constraints, such as restricting to Gaussian input states or fixing the photon-number fluctuation, are required to operate in a physically allowed regime.

Another important scenario is displacement sensing with a generator of $\hat{H}=\hat{p}$. First considering a single probe, the Fisher information reads $
I_F[\hat{U}\left(\alpha\right)\ket{\psi}]/4
={\rm var} \left(\hat{p} \right)_{\psi}.
$ 
To achieve the optimal precision, we need to maximize ${\rm var} \left(\hat{p} \right)_{\psi}$, subject to the energy constraint $N_S=\expval{\hat{a}^\dagger \hat{a}}\equiv \left({\rm var} \left(\hat{p} \right)_{\psi}+{\rm var} \left(\hat{q} \right)_{\psi}\right)/4-1/2$. Here, zero-mean input states are chosen because displacement does not enhance the precision. Taking into account the uncertainty principle, ${\rm var} \left(\hat{p} \right)_{\psi}\cdot{\rm var} \left(\hat{q}\right)_{\psi}\ge 1$, and the energy constraint, it shows that ${\rm var} \left(\hat{p} \right)_{\psi}$ is maximized when the input is in a squeezed-vacuum state, with the optimal Fisher information
\be 
I_F[\hat{U}\left(\alpha\right)\ket{\psi}]/4=\left(\sqrt{N_S}+\sqrt{N_S+1}\right)^2.
\label{F_single_mode}
\ee 
We now consider $M$ probes with a total energy constraint $N_S$. For separable input states, the Fisher information is additive (Eq.~\eqref{F_varU_product}). Since the single-mode Fisher information in Eq.~\eqref{F_single_mode} is concave with respect to the energy $N_S$, the Fisher information for $M$ separable probes becomes
\be 
I_F\le I_F^{\rm max,C}\equiv  4M\left(\sqrt{N_S/M}+\sqrt{N_S/M+1}\right)^2\simeq 16 N_S,
\label{F_varU_product_CV}
\ee 
whose maximum is achieved with the input being in an iid product of single-mode squeezed vacuum state. Asymptotically, it is independent of the number of probes $M$.

With an entangled input state, we need to consider the overall Fisher information of Eq.~\eqref{F_var_u}:
\be
I_F[\hat{U}^{\otimes M}\left(\alpha\right)\ket{\bm \psi_M}]/4
= M \cdot {\rm var} \left(\hat{P} \right)_{\bm \psi_M},
\label{F_var_u_CV}
\ee
where $\hat{P}=\sum_\ell \hat{p}_\ell/\sqrt{M}$ is defined as a global momentum quadrature, with the corresponding global position quadrature defined as $\hat{Q}=\sum_\ell \hat{q}_\ell/\sqrt{M}$. Because $\hat{P}$ and $\hat{Q}$ satisfy the canonical commutation relation and can be obtained by applying a passive linear transform on the original single-mode quadrature operators, they can be regarded as being effectively produced by sending the original input modes through a balanced beam-splitter array. As energy is preserved under a beam-splitter transform, it is clear that concentrating all the energy on the global mode of $\hat{P}$ and $\hat{Q}$ is the optimal, as the other global modes are all in vacuum states. At this juncture, we have reduced the multimode problem to a single-mode problem, and the Fisher information 
\be
I_F\le I_F^{\rm max,E}
\equiv 4M \left(\sqrt{N_S}+\sqrt{N_S+1}\right)^2\simeq 16M N_S
\label{F_var_u_CV_optimal}
\ee 
is achieved by an entangled input state whose corresponding global mode is in a single-mode squeezed-vacuum state. This sub-SQL performance is achieved in a special DQS protocol, as explained in Sec.~\ref{sec:CV_distributed_sensing}.

\subsection{Precision limits in the presence of noise}

The above analyses assume pure states so that the Fisher information has a simple form represented in Eqs.~\eqref{F_var} and~\eqref{F_var_u}. However, experimental imperfections are in general inevitable, rendering the output state mixed and the evaluation of the Fisher information more challenging. Although an exact optimization of the Fisher information is difficult, one can obtain upper bounds on the Fisher information~\cite{escher2011}. One can also focus on certain specific classes of input states, such as the Gaussian states~\cite{Weedbrook2012}, to make the problem tractable.

Here, we consider the CV example of displacement estimation introduced in Sec.~\ref{sec:CV_Fisher} and explain the above two approaches. We consider the most common noise in photonic sensing---loss. Suppose pure loss occurs before the probing process, the input-output relation for a single mode is
\be 
\hat{\rho}(\alpha)= \hat{U}(\alpha)[\calN_\eta(\hat{\rho})]\hat{U}^\dagger(\alpha),
\label{rho_alpha_loss}
\ee 
which augments a lossy channel into the sensing model described Eq.~\eqref{rho_alpha}. Similar to Eq.~\eqref{rho_output}, the output $M$-mode state is
\be 
\hat{\bp}_M\left(\alpha\right)=\hat{U}^{\otimes M}\left(\alpha\right)[\calN^{\otimes M}_\eta\left(\hat{\bp}_{M}\right)]\hat{U}^{\dagger\otimes M}\left(\alpha\right).
\label{rho_alpha_loss_M}
\ee 

\subsubsection{General upper bounds}
\label{general_bound}
Several general upper bounds for the DQS and DCS protocols will be presented (details in Ref.~\cite{xia2018repeater}). The probe state after loss is in general mixed, expressed as $\hat{\bp}_{M}=\sum_\ell c_\ell \ket{\psi_\ell}\bra{\psi_\ell}$, where $\sum_\ell c_\ell=1$ are positive eigenvalues and $\psi_\ell$'s are the eigenstates. Per convexity of the Fisher information, we have
\be
I_F[\hat{\bp}_M(\alpha)]\le \sum_\ell c_\ell I_F[\hat{U}\left(\alpha\right)^{\otimes M}\ket{\psi_\ell}].
\label{F_convex}
\ee 
For each pure state, similar to Eq.~\eqref{F_var_u}, we can explicitly express
$
I_F[\hat{U}\left(\alpha\right)^{\otimes M}\ket{\psi_\ell}]
=4 {\rm var} \left(\sum_k {\hat{p}_k^\prime} \right)_{\psi_\ell},
$
where ${\hat{p}_k^\prime}$ is the momentum quadrature of the $k$th probe after the lossy channel.
Combining the above with Eq.~\eqref{F_convex}, we have
\be
I_F[\hat{\bp}_M(\alpha)]/4
\le
\expval{\left(\sum_k {\hat{p}_k^\prime} \right)^2}_{\hat{\bp}_M(\alpha)}-
\sum_\ell c_\ell \expval{\left(\sum_k {\hat{p}_k^\prime} \right)}^2_{\psi_\ell}
\le
{\rm var}\left(\sum_k {\hat{p}_k^\prime} \right)_{\hat{\bp}_M(0)},
\ee
where the inequality $\sum_\ell c_\ell x_\ell^2\ge \left(\sum_\ell c_\ell x_\ell\right)^2$ and the invariance of quadrature variances under displacement are used. 
Because the pure-loss channel transforms $\hat{p}_k^\prime=\sqrt{\eta}\hat{p}_k+\sqrt{1-\eta} \hat{p}_{e_k}$, where $\hat{p}_{e_k}$ and $\hat{p}_k$ are the momentum quadratures of the environment and the mode before the loss, we immediately have
\be
I_F[\hat{\bp}_M(\alpha)]/4 \le  \eta {\rm var}\left(\sum_k {\hat{p}_k} \right)+\left(1-\eta\right)M,
\ee 
where the variance is evaluated before the pure-loss channel. Combining with the analyses in Eq.~\eqref{F_var_u_CV_optimal}, an upper bound (UB) for the Fisher information is derived as
\be
I_F\le I_F^{\rm UB,E}\equiv  \eta I_F^{\rm max,E}+4\left(1-\eta\right)M.
\label{F_var_u_CV_optimal_UB_loss}
\ee 
At $\eta=1$, the UB recovers Eq.~\eqref{F_var_u_CV_optimal} and is therefore achievable. However, it is unclear whether Eq.~\eqref{F_var_u_CV_optimal_UB_loss} is achievable at $\eta < 1$. One also finds that the method utilizing purification to reduce to pure states calculations~\cite{escher2011} gives the same UB as Eq.~\eqref{F_var_u_CV_optimal_UB_loss}.

Similarly, the Fisher information UB in DCS is derived as
\be
I_F\le I_F^{\rm UB,C}\equiv  \eta I_F^{\rm max,C}+4\left(1-\eta\right)M,
\label{F_varU_product_CV_UB_loss}
\ee 
which recovers Eq.~\eqref{F_varU_product_CV} at $\eta=1$, while its achievability remains unclear for $\eta < 1$.

\subsubsection{Optimal Gaussian states}
\label{opt_Gaussian}

Now we discuss the optimal Gaussian states for the DQS and DCS protocols.~\cite{zhuang2018distributed}. In the following, we first show that a single-mode squeezed-vacuum state is the optimal Gaussian probe for single-mode sensing. 

The Gaussian input state $\hat{\rho}$ in Eq.~\eqref{rho_alpha_loss} is fully characterized~\cite{Weedbrook2012} by its quadratures' mean vector ${\boldsymbol a}$ and covariance matrix ${\bf V}$, where the quadratures are defined as ${\rm Re}(\hat{a})$ and ${\rm Im}(\hat{a})$.  Then, writing $\hat{\rho}$ as the Gaussian state $\hat{\rho}_G({\boldsymbol a},{\bf V})$, we get $\hat{\rho}_G(\sqrt{\eta}\,{\boldsymbol a} + {\boldsymbol \alpha}, \eta{\bf V} + (1-\eta){\bf I}/4)$ for a Gaussian state $\hat{\rho}(\alpha)$, where ${\boldsymbol \alpha} \equiv [\alpha,0]$ and ${\bf I}$ is the $2\times 2$ identity matrix. Thus,
$\hat{\rho}_G(\sqrt{\eta}\,{\boldsymbol a} + {\boldsymbol \alpha} + {\boldsymbol \epsilon}, \eta{\bf V} + (1-\eta){\bf I}/4)$ for the Gaussian state $\hat{\rho}(\alpha+ \epsilon)$, where ${\boldsymbol \epsilon} \equiv [\epsilon,0]$.  The quadrature covariance matrix of an arbitrary $\hat{\rho}_G({\boldsymbol a},{\bf V})$ can always be written in the form of ${\bf V} = {\bf R}_\theta {\bf V}_{\rm diag}{\bf R}_\theta^T$, where ${\bf V}_{\rm diag} =  {\rm Diag}[(2n+1)e^{-r}/4,(2n+1)e^{r}/4]$ with $r\ge 0$, $n\ge 0$, and 
\be
{\bf R}_\theta = \left[\begin{array}{cc}
\cos(\theta) & \sin(\theta) \\
-\sin(\theta) & \cos(\theta) \end{array}
\right].
\ee
With this ${\bf V}$ representation, Ref.~\cite{scutaru1998fidelity} shows how to evaluate the Uhlmann fidelity between $\hat{\rho}_G(\sqrt{\eta}\,{\boldsymbol a} + {\boldsymbol \alpha},\eta {\bf V} + (1-\eta){\bf I}/4)$ and $\hat{\rho}_G(\sqrt{\eta}\,{\boldsymbol a} + {\boldsymbol \alpha} + {\boldsymbol \epsilon}, \eta{\bf V} + (1-\eta){\bf I}/4)$. Using the result of that evaluation in Eq.~(\ref{F_Uhlmann}) gives
\begin{equation}
I_F[\hat{\rho}(\alpha)] = 
\frac{4\{e^{r}(1-\eta) + (2n+1)\eta[e^{2r}\cos^2(\theta) + \sin^2(\theta)]\}}{(e^{r}(1-\eta)+(2n+1)\eta)[(2n+1)\eta e^{r} + 1-\eta]}.
\end{equation}
This expression's maximum over $\theta$ and $n$ occurs when $\theta = n = 0$, in which case $\max_{\theta,n}I_F[\hat{\rho}(\alpha)]=4 /(\eta e^{-r}+ 1-\eta)$ with mean the photon-number constraint $N_S = {\boldsymbol a}^T{\boldsymbol a} + [\cosh(r) - 1]/2$.  From this result it is clear that ${\boldsymbol a} = {\bf 0}$ is optimal, so the UB
\be
I_F^{\rm Gauss} \le I_F^{\rm Gauss, max}
\equiv  \left(\frac{\eta}{4(\sqrt{N_S+1} + \sqrt{N_S})^2} + \frac{1-\eta}{4}\right)^{-1},
\ee
is achieved by a single-mode squeeze-vacuum state.

One can then solve the separable Gaussian input case, via considering the power distribution $\{N_m\}_{m=1}^M$ across different modes. Optimization over the additive Fisher information gives the UB
\be
I_F^{\rm Gauss,C} \le I_F^{\rm Gauss, C, max}\equiv \left(\frac{\eta}{M\left(\sqrt{N_S/M+1}+\sqrt{N_S/M}\right)^2}+\frac{1-\eta}{M}\right)^{-1},
\label{F_varU_product_CV_loss}
\ee 
which is achievable by an iid product of single-mode squeezed-vacuum state. At $\eta=1$, Eq.~\eqref{F_varU_product_CV_loss} reduces to Eq.~\eqref{F_varU_product_CV}, confirming that in a lossless case the Gaussian state is the optimum.
For the entangled input, as we explained in Sec.~\ref{sec:CV_distributed_sensing}, one can reduce the $M$-probing problem to a single probing and obtain
\be 
I_F^{\rm Gauss,E} \le I_F^{\rm Gauss, E, max}\equiv \left(\frac{\eta}{M\left(\sqrt{N_S+1}+\sqrt{N_S}\right)^2}+\frac{1-\eta}{M}\right)^{-1},
\label{F_var_u_CV_optimal_loss}
\ee 
which is achievable by an entangled input designed in Sec.~\ref{sec:CV_distributed_sensing}. Similarly, at $\eta=1$, Eq.~\eqref{F_var_u_CV_optimal_loss} reduces to Eq.~\eqref{F_var_u_CV_optimal}, affirming that in the lossless case the Gaussian state is optimal.

\subsection{Multiparameter quantum Cram\'{e}r-Rao bound}
\label{sec:CR_multi}

In distributed sensing, although a single scalar function is being estimated, one in general needs a multi-parameter quantum CR bound to analyze the performance limit of unbiased estimation. Here, we generalize the analyses in Sec.~\ref{sec:single_parameter_CR}. 

Consider the estimation of parameters $\bm \alpha$ of the state $\hat{\bp}_M\left(\bm \alpha\right)=\hat{U}\left(\bm \alpha\right)\hat{\bp}_M\hat{U}^\dagger\left(\bm \alpha\right)$ given in Eq.~\eqref{rho_output_general}. For distributed sensing, it suffices to consider $\hat{U}\left(\bm \alpha\right)=\exp\left(-\sum_{m=1}^M \hat{H}_m \alpha_m\right)$, with generators $\hat{H}_m$ commuting. In a general multi-parameter estimation scenario, one performs measurement and obtains an estimator $\tilde{\bm \alpha}=\left(\tilde{\alpha}_1,\cdots, \tilde{\alpha}_M\right)$. The precision is characterized by the covariance matrix with elements
\be 
\bm V_{mn}=\expval{\left(\tilde{\alpha}_m-\alpha_m\right)\left(\tilde{\alpha}_n-\alpha_n\right)}.
\ee 
Similar to Eq.~\eqref{LSD}, we introduce the SLD $\hat{L}_{\alpha_m}$ for each parameter. Then the Fisher information matrix has elements
\be
\bm H_{mn}=\tr\left[\hat{\bp}_M\left(\bm \alpha\right)\frac{\hat{L}_{\alpha_m}\hat{L}_{\alpha_n}+\hat{L}_{\alpha_n}\hat{L}_{\alpha_m}}{2}\right].
\label{Fisher_mat}
\ee 
For arbitrary weight matrix $\bm G$, we have the CR bound
\be
\Tr\left[\bm G \bm V\right]\ge \Tr\left[\bm G \bm H^{-1}\right].
\label{CR_multi}
\ee 

For distributed sensing, we focus on the estimation of $\bar{\alpha}\equiv \sum_{\ell=1}^M w_\ell \alpha_\ell$ in Eq.~\eqref{weighted_average} and therefore one can choose $\bm G=\bm w^T \bm w$ in Eq.~\eqref{CR_multi} to obtain the ultimate limit of distributed sensing in estimating $\bar{\alpha}$ as
\be 
\delta \alpha^2\equiv \tr\left[\bm w^T \bm w \bm V\right]\ge  \tr\left[\bm w^T \bm w \bm H^{-1}\right].
\label{CR_distributed}
\ee 
An alternative bound from the right logarithmic derivative can be obtained similarly. For multiple-parameter estimation, the readers can refer to Refs.~\cite{Genoni_2013,steinlechner2013quantum,ast2016reduction,Baumgratz_2016,nair2018quantum,zhuang2017entanglement}. 

Evaluating the Fisher information matrix in Eq.~\eqref{Fisher_mat} is in general challenging, but can be done for Gaussian states~\cite{oh2020optimal}.

\section{Applications of Distributed Quantum Sensing}
\label{sec:applications}
DQS has been recently demonstrated in two experiments based on CVs. We next introduce the two experimental setups and discuss their applications in optical phase and RF sensing. 

\subsection{Optical phase sensing}
\begin{figure}[bth]
    \centering
    \includegraphics[width=1\textwidth]{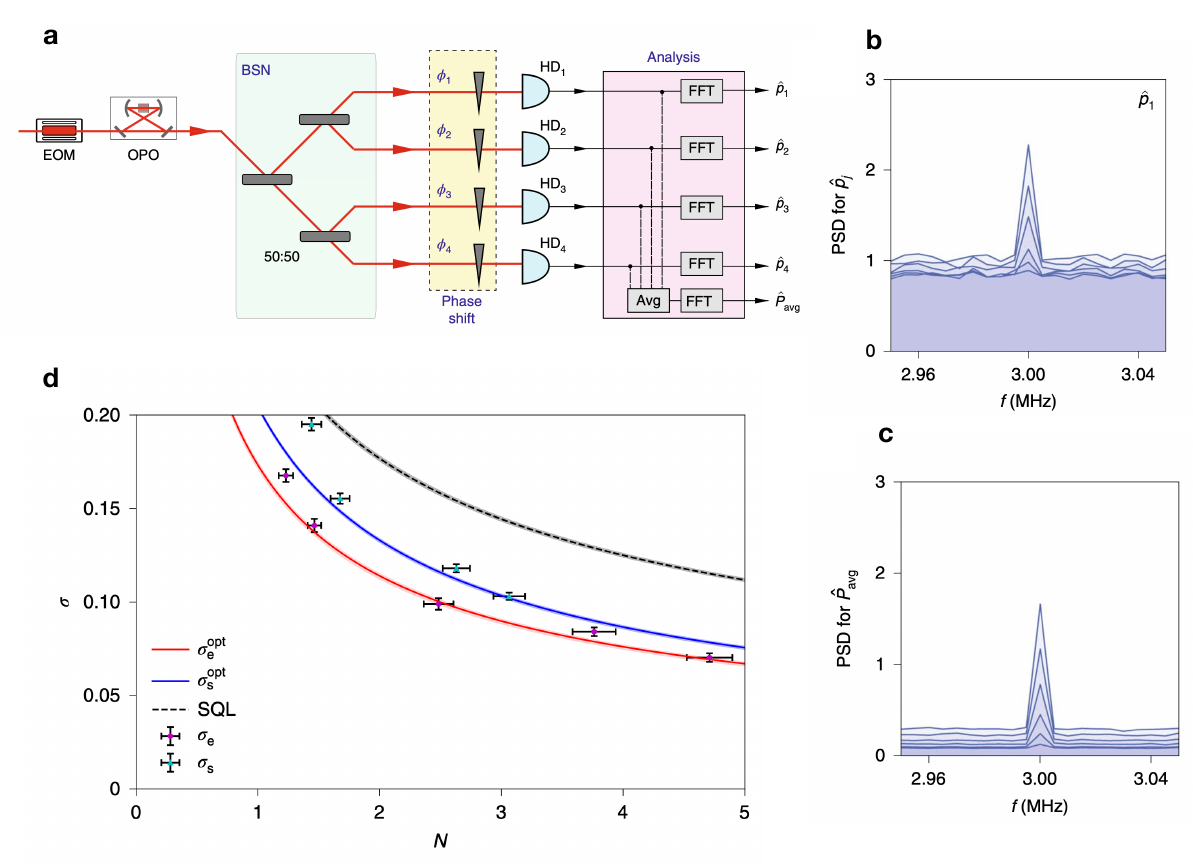}
    \caption{DQS experiment for optical phase sensing using CV multipartite entanglement. (a) Experimental diagram. OPO: optical parametric oscillator; EOM: electro-optic modulator; BSN: beam-splitter network; HD: homodyne detector; FFT: fast Fourier transform. (b) The frequency-domain measurement data acquired by the first sensor, showing a signal peak at 3 MHz and a noise floor. (c) The averaged measurement data in the frequency domain. The averaged data show a higher signal-to-noise ratio than the measurement data acquired at a single sensor due to the entanglement-enabled cancellation of measurement noise from different sensors. (d) Comparison of the measurement sensitivities of the DQS scheme and the DCS scheme. $\sigma_e^{\rm opt}$: optimal measurement sensitivity for DQS; $\sigma_s^{\rm opt}$: optimal measurement sensitivty for DCS; $\sigma_e$: experimental measurement sensitivity for DQS; $\sigma_s$: experimental measurement sensitivity for DCS; SQL: standard quantum limit. Figure reproduced from~\cite{guo2020distributed}.}
    \label{fig:phase_sensing}
\end{figure}
Guo {\em et al.} verified the principle of DQS in a CV quantum-optics platform. Fig.~\ref{fig:phase_sensing}(a) illustrates the experimental setup. An electro-optic modulator (EOM) creates 3 MHz sidebands of a 1550-nm beam, which is subsequently injected into an optical parametric oscillator (OPO) cavity as the seed. The OPO cavity is pumped below the oscillation threshold and produces displaced squeezed light at the 3 MHz sidebands at its output. The displaced squeezed light is then split into four arms by a beam-splitter network (BSN) comprising three 50:50 beam splitters. Each arm serves as a sensor node to probe an optical phase shift controlled by wave plates. The objective of the sensing task is to measure a global parameter, the average phase shift, across the sensor network. To this end, a homodyne detector (HD) measures the quadrature displacement at each sensor. The measurement data from four sensors are postprocessed to derive the average phase shift. The frequency-domain data acquired by a single sensor are plotted in Fig.~\ref{fig:phase_sensing}(b), where the signal peak at 3 MHz and the measurement-noise floor can be clearly identified. The spectrum of the signal averaged over four sensors is depicted in Fig.~\ref{fig:phase_sensing}(c). Due to the quantum correlation between the measurement noise at different sensors, the noise power after averaging is significantly reduced, leading to a higher signal-to-noise ratio than that at a single sensor. In Fig.~\ref{fig:phase_sensing}(d), The performance of DQS (red solid curve) is compared with that of DCS (blue solid curve) and the SQL (black dashed curve). The DCS experiment employs separable squeezed light at the sensors subject to the same total mean photon number as the DQS experiment based on entangled probe states, i.e., adopting the first resource counting scheme discussed at the outset. The derivation of the SQL assumes that the sensors utilize classical coherent light to estimate the average optical phase shift. DQS outperforms both the DCS and SQL at all mean photon-number levels. The DCS performs better than the SQL due to single-mode squeezed light used by each sensor.

\subsection{RF sensing}
\label{sec:rf}

\begin{figure}[bth]
    \centering
    \includegraphics[width=1\textwidth]{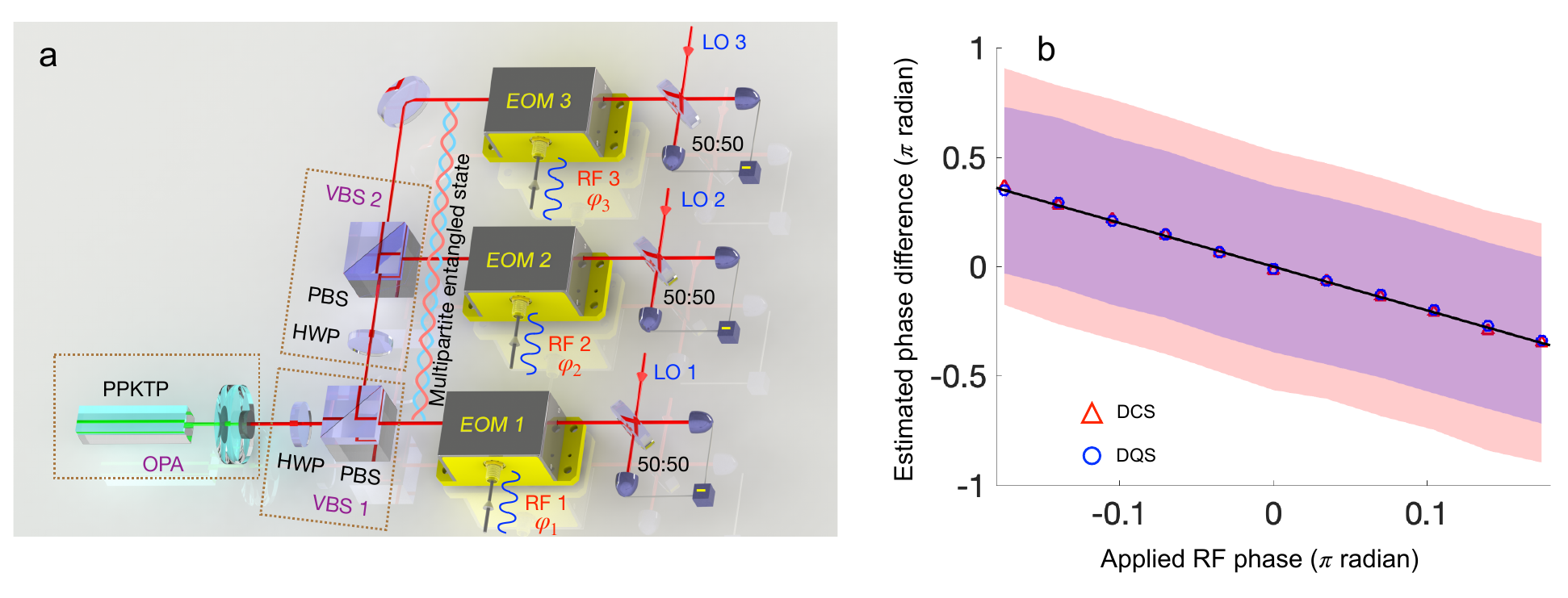}
    \caption{DQS experiment for RF sensing using CV multipartite entanglement. (a) The experimental diagram. PPKTP: periodically-poled KTiOPO$_4$; OPA: optical parametric amplifier; VBS: variable beam splitter; HWP: half-wave plate; PBS: polarizing beam splitter; EOM: electro-optic modulator; LO: local oscillator. (b) Estimation of RF-field phase difference at an edge sensor node by DQS (blue) and DCS (red). The estimation uncertainty (shades) of DQS clearly outperforms that of DCS. Figure reproduced from~\cite{xia2020demonstration}.}
    \label{fig:RF_sensing}
\end{figure}

In addition to optical phase sensing, DQS can enhance a wide range of sensing applications by introducing transducers that covert the probed physical parameters into modulations on the entangled probe state in the optical domain. Xia {\em et al.} recently reported an experiment of DQS for RF sensing, as sketched in Fig.~\ref{fig:RF_sensing}(a). Squeezed light is produced from an optical parametric amplifier (OPA) cavity in which a type-0 periodically-poled KTiOPO$_4$ (PPKTP) crystal is embedded. The OPA cavity is seeded with 1550-nm light such that its output consists of a strong classical coherent state at the central wavelength and phase squeezed vacuum state residing at the sidebands. Two variable beam splitters (VBSs), each composed of a half-wave plat (HWP) and a polarizing beam splitter (PBS), constitute a quantum circuit that configures the multipartite entangled state. As we discuss below, one needs to tailor the structure of the multipartite entangled state to achieve the optimal performance for a specified RF-sensing task such as average RF-field amplitude estimation or angle-of-arrival (AoA) estimation via RF-field phase difference measurements. In the experiment, a three-mode entangled state is generated for probing a global property of the RF field observed at three RF-photonic sensors. The RF field at the $m$th RF-photonic sensor can be represented as
\begin{equation}
    \mathcal{E}_m(t) = E_m\cos(\omega_c t +\varphi_m),
\end{equation}
where $A_m$ is the RF amplitude, $\omega_c$ is the RF carrier frequency, $t$ is time, and $\varphi_m$ is the RF phase. Under the weak-field limit, an EOM converts the RF field into a displacement
\begin{equation}
    \alpha_m \propto E_m \varphi_m
\end{equation}
on the phase quadrature. A balanced homodyne detector measures the quadrature displacement, yielding an estimator $\tilde{\alpha}_m$. Postprocessing on the $\tilde{\alpha}_m$'s from all three sensors leads to an estimation on the RF-field property of interest.

\begin{figure}[bth]
    \centering
    \includegraphics[width=1\textwidth]{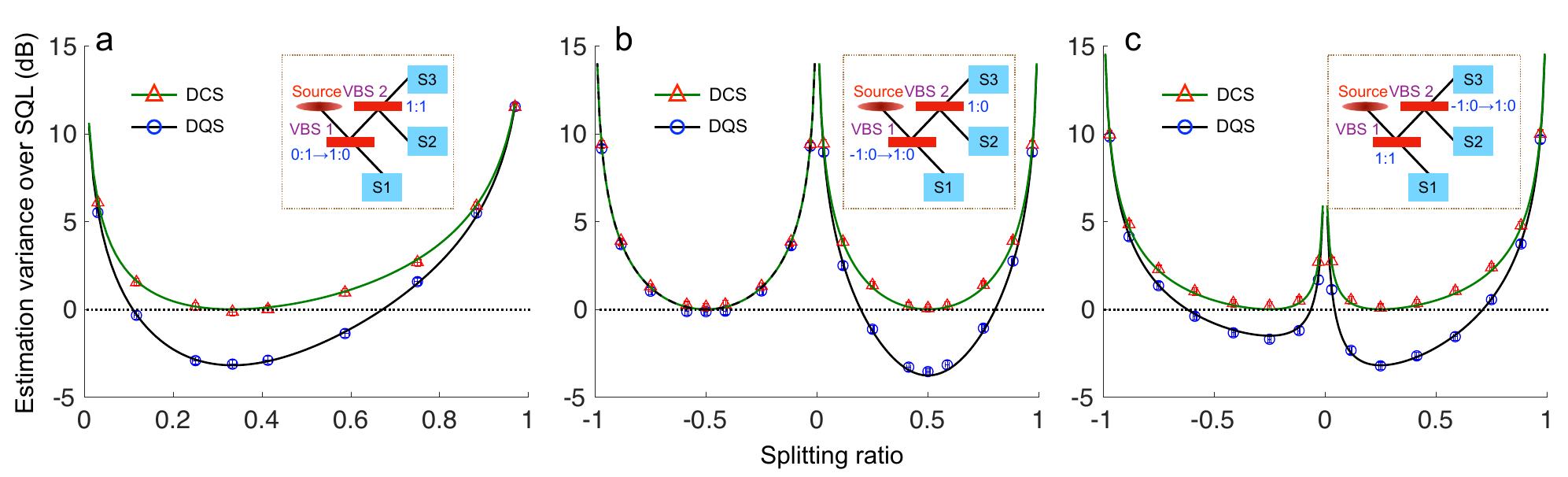}
    \caption{Optimization of multipartite entangled state for (a) RF-field average amplitude estimation, (b) phase-difference estimation at a central sensor node, and (c) phase-difference estimation at an edge sensor node. Insets show the tuning ranges of the splitting ratios for VBSs. Circles: DQS experimental data; triangles: DCS experimental data; solid curves: theory. Figure reproduced from~\cite{xia2020demonstration}.}
    \label{fig:entanglement_optim}
\end{figure}

Three RF-sensing tasks were investigated in the experiment: RF-field average amplitude estimation, phase-difference estimation at a central sensor node, and phase-difference estimation at an edge sensor node. The same RF-sensing tasks were carried out by DCS as a performance benchmark. In DCS that defines the SQL, the entangled state at the sidebands was turned off while the strong coherent state at the central wavelength was retained, in line with the second resource-counting scheme discussed in Sec.~\ref{sec:overview}. Fig.~\ref{fig:RF_sensing}(b) compares the estimation uncertainty of DQS (blue shade) and DCS (red shade) under the task of measuring the phase difference at an edge node. DQS enjoys an appreciable advantage over DCS by virtue of the multipartite entangled state shared by the three sensors.

The entangled senor network is reconfigurable in that the multipartite entangled state can be tailored by the quantum circuit to minimize the estimation uncertainty for a specified RF-sensing problem. To build a connection between the structure of the multipartite entangled state and the performance of a RF-sensing problem, Xia {\em et al.} optimized the multipartite entangled state by tuning the parameters of the quantum circuit and measured the variance of estimation under different settings~\cite{xia2020demonstration}. The theoretical predictions (solid curves) and experimental data for DQS (circles) and DCS (triangles) are depicted in Fig.~\ref{fig:entanglement_optim}. Entanglement optimization for three RF-sensing problems are studied: RF-field average amplitude estimation (Fig.~\ref{fig:entanglement_optim}(a)), phase-difference estimation at a central sensor node (Fig.~\ref{fig:entanglement_optim}(b)), and phase-difference estimation at an edge sensor node (Fig.~\ref{fig:entanglement_optim}(c)). In all cases, the splitting ratio of one VBS of the quantum circuit is tuned while the other VBS's splitting ratio is fixed. In the optimization of the multipartite entangled state for phase-difference estimation, the phase of one arm of the quantum state is flipped between 0 and $\pi$ as an additional knob. The largest advantage of DQS over DCS is only achieved at an appropriately chosen beam-splitter ratio and phase, highlighting the necessity of tailoring the entangled probe state for different RF-sensing tasks. Notable, a consequence of the quantum correlations between the measurement noise at different sensors, the DQS estimation variances show asymmetric behaviors as the phase is flipped~\cite{xia2020demonstration}.

\subsection{Machine-learning applications}
Quantum machine learning~\cite{schuld2019quantum,lloyd2013quantum,wichert2013principles,dunjko2016quantum,biamonte2017quantum,dunjko2018machine,faber2018alchemical} has recently spurred broad interest as it creates new opportunities to harness the power of quantum resources to reduce the complexity of many data-processing problems. At present, however, large-scale fault-tolerant random access quantum memories and NISQ devices with sufficient circuit depths remain elusive to allow many quantum machine-learning schemes to achieve an advantage over classical machine-learning schemes. 

\begin{figure}[bth]
    \centering
    \includegraphics[width=0.6\textwidth]{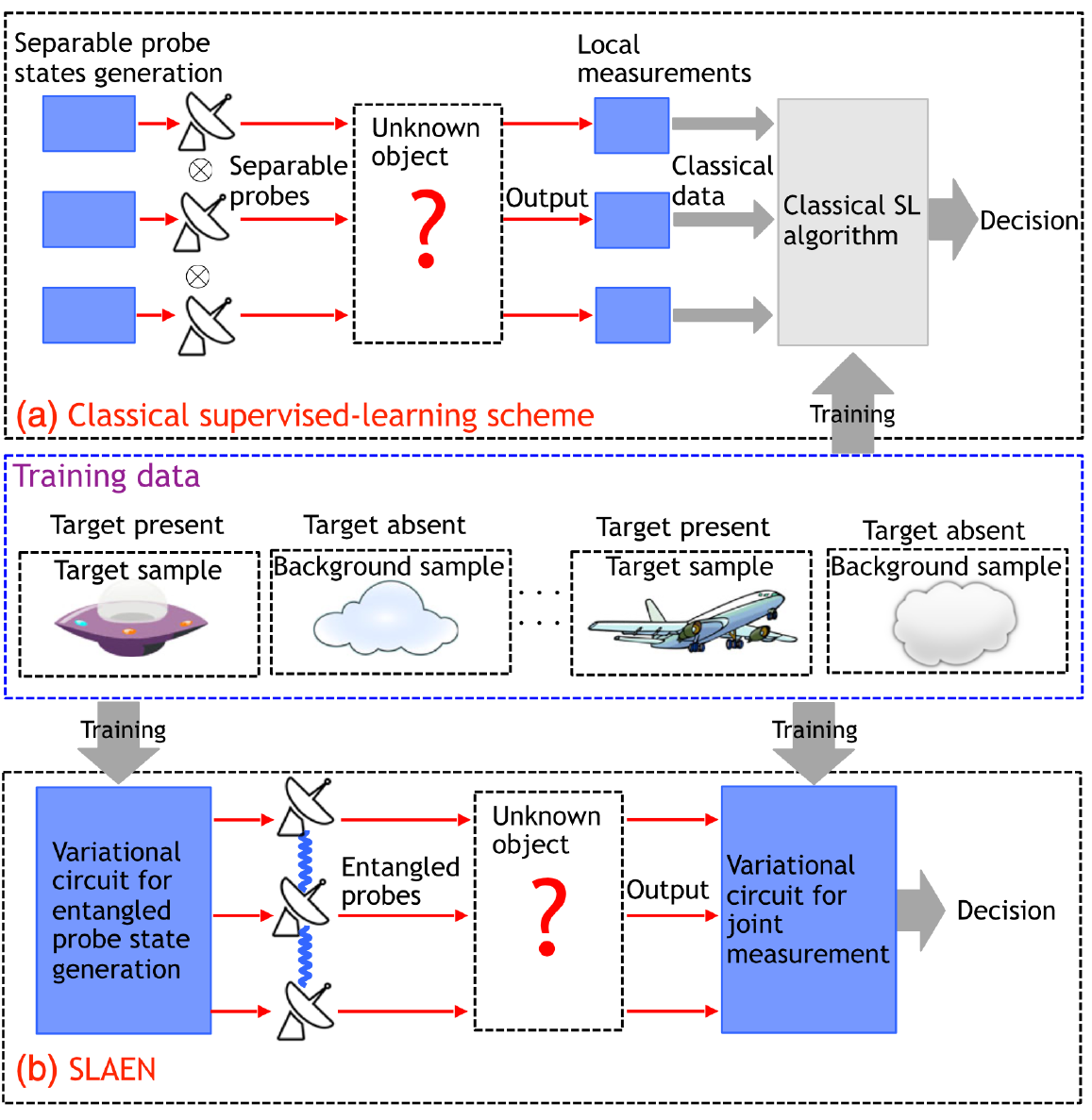}
    \caption{(a) Classical supervised-learning (SL) scheme. (b) Supervised-learning assisted by an entangled sensor network (SLAEN). Both schemes rely on training data to configure the classical algorithms and variational quantum circuits. Figure reproduced from Ref.~\cite{zhuang2019physical}.
    \label{fig:SLAEN}
    }
\end{figure}

Many real-world applications rely on sensors to acquire data for classical machine-learning algorithms to perform further processing, e.g., classification or compression. A conventional classical supervised-learning scheme is illustrated in Fig.~\ref{fig:SLAEN}(a) using a target classification example. Sensors in separable quantum states probe an object, generating local measurement data that are processed by a classical supervised-learning algorithm. The classical supervised-learning algorithm is first trained using known objects such that it becomes capable of processing data acquired from interrogating unknown objects once the training completes. The involvement of sensors opens a window to leverage DQS to achieve quantum-enhanced performance in data processing. In particular, as machine-learning applications often only requires the acquisition of a global property, such as the class of the sample, it is natural that DQS can be utilized to benefit. Supervised learning assisted by an entangled sensor network (SLAEN)~\cite{zhuang2019physical}, illustrated in Fig.~\ref{fig:SLAEN}(b), is a recently developed hybrid quantum-classical machine-learning scheme that reaps such benefits from DQS. In SLAEN, a variational quantum circuit is trained by a classical supervised-learning algorithm to generate entangled states shared by the sensors. After interrogating the object, a second variational quantum circuit processes the quantum state prior to measurements. The measurement data are fed to a classical algorithm to complete data processing. Like the classical supervised-learning scheme, SLAEN also rest upon training data to configure the variational quantum circuit and the classical algorithm to achieve the optimal performance and largest quantum advantage. 

\begin{figure}[bth]
    \centering
    \includegraphics[width=0.8\textwidth]{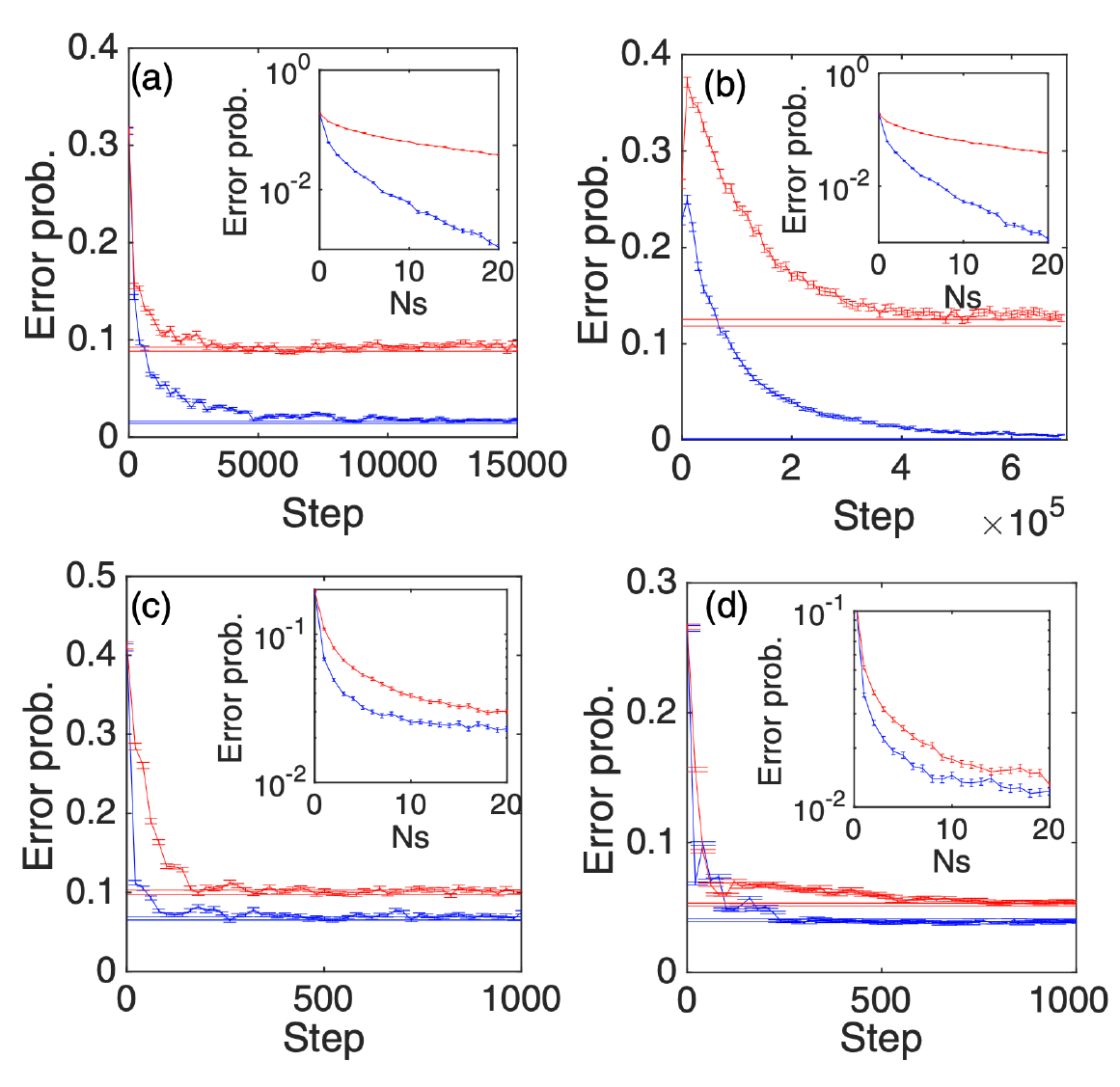}
    \caption{Error probability vs training steps for SLAEN (blue curves) and classical supervised learning (red curves). (a), (b), (c), and (d) assume different source brightness and system efficiencies in their simulations (see ~\cite{zhuang2019physical} for more details). Insets: the scaling of the error probability with respect to the source brightness. Figure reproduced from Ref.~\cite{zhuang2019physical}.}
    \label{fig:SLAEN_training}
\end{figure}

The error probabilities of SLAEN (blue curves) and the classical supervised-learning scheme (red curves) during the training process are plotted in Fig.~\ref{fig:SLAEN_training} for different physical parameters. It is evident that SLAEN achieves a quantum advantage over the classical supervised-learning scheme by enabling a reduced error probability in solving data-classification problems. Since sensors are ubiquitous for data collection, SLAEN and DQS would pave a distinct and broad route for quantum-enhanced data processing in the NISQ era, as demonstrated in a recent experiment for the classification of RF signals~\cite{xia2020quantum}.

\section{Outlook}
\label{sec:outlook}
As a new paradigm for quantum metrology, DQS endows new quantum-enhanced capabilities for applications built upon a network of sensors. Yet, a few fundamental questions about DQS remain unanswered. First, although Sec.~\ref{sec:distributed_displacement} has elucidated the optimal DQS scheme for displacement sensing in the absence of loss and the optimal Gaussian scheme in the presence of loss, and Sec.~\ref{variation:phase} provided the optimal Gaussian DQS scheme for phase sensing, a general optimal DQS scheme remains unknown and warrants future investigations. Second, it is well known that non-Gaussian resources such as the GKP state~\cite{gottesman2001encoding} lead to universality in CV quantum computing~\cite{lloyd1999quantum}. It is shown that the GKP state can improve the performance of DQS by compensating for loss~\cite{zhuang2020distributed}, but it calls for further studies to assess the utility of non-Gaussian probe states and measurements in DQS. Third, DV DQS protocols are useful for a variety of applications such as magnetic and stress sensing, but they are less explored than CV DQS protocols. In particular, it would be worthwhile to conduct a systematic analysis on the performance of DV DQS protocols subject to practical nonidealities. Also, scalable creation of DV entangled states and distributing them to DV sensors need additional research.

From an application perspective, DQS can be exploited to address sensing and data-processing problems in different physical domains. For example, a recent theoretical proposal envisages use of DQS to boost the performance of optical gyroscopes for navigation applications~\cite{grace2020quantum}. Moreover, DQS could be leveraged to enhance the readout signal-to-noise ratio of an array of optomechanical transducers used in inertial sensing~\cite{hines2020optomechanical} and atomic force microscopes~\cite{sugimoto2007chemical}. It would also be foreseeable that DQS would benefit biological measurements for, e.g., molecular identification and tracking~\cite{taylor2013biological,luan2018silicon}.

The experimental demonstrations of DQS so-far are all based on table-top quantum-optics platforms. Recent advances in on-chip generation of squeezed light~\cite{dutt2015chip,zhao2020near,vaidya2019broadband,cernansky2019nanophotonic} and entangled photons~\cite{ramelow2015silicon,wang2018multidimensional}, in conjunction with efficient on-chip transducers~\cite{jiang2020efficient}, have promised a more scalable and cost-effective solution for the widespread deployment of CV DQS protocols. On the DV front, large-scale integration of color centers with photonic circuits has been recently achieved~\cite{wan2020large}, opening a route for utilizing DQS to benefit chip-scale magnetic and electric-field sensing applications.

\section*{Acknowledgment}
Z.Z. are grateful for support from Office of Naval Research (ONR) Grant No.~N00014-19-1-2190 and National Science Foundation (NSF) Grant No.~ECCS-1920742 and CCF-1907918. Q.Z. acknowledges funding from Defense Advanced Research Projects Agency (DARPA) under Young Faculty Award (YFA) Grant No.~N660012014029 and University of Arizona.

\bibliographystyle{iopart-num}
\bibliography{myref}

\end{document}